\newif\ifconfver
\confverfalse      
\confvertrue        

\newif\ifplainver  
\plainvertrue
\plainverfalse

\newif\ifhide  
\hidetrue

\ifplainver
    \confverfalse   
\fi

\ifconfver
     \documentclass[10pt,twocolumn,twoside]{IEEEtran}
\else
    \ifplainver
        \documentclass[11pt]{article}
        \usepackage{fullpage}
    \else
        \documentclass[12pt,draftcls,onecolumn]{IEEEtran}
    \fi
\fi

\usepackage{etoolbox}%
\usepackage{xpatch}
\usepackage{blindtext}
\usepackage{setspace}

\usepackage{tocloft}%
\newlength{\articlesectionshift}%
\let\LaTeXStandardSection\section
\let\LaTeXStandardTheSection\thesection
\let\LaTeXStandardTheSubSection\thesubsection
\let\LaTeXStandardTheSubSubSection\thesubsubsection
\let\LaTeXStandardTheParagraph\theparagraph

\makeatletter
\newcounter{titlecounter}

\xpretocmd{\maketitle}{\ifnumgreater{\value{titlecounter}}{1}}{\clearpage}{}{} 
\xpatchcmd{\maketitle}{\let\maketitle\relax\let\@maketitle\relax}{\refstepcounter{titlecounter}\begingroup
  \addtocontents{toc}{\begingroup\addtolength{\cftsecindent}{-\articlesectionshift}}%
  \addcontentsline{toc}{section}{\protect{\numberline{\thetitlecounter}{\@title~ \@author}}}%
  \addtocontents{toc}{\endgroup}
}{%
  \typeout{Patching was successful}
}{%
  \typeout{patching failed}
}%

\def\@IEEEdestroythesectionargument#1{\LaTeXStandardSection{#1}}%

\xapptocmd{\maketitle}{%
\renewcommand{\thesection}{\LaTeXStandardTheSection}%
\renewcommand{\thesubsection}{\LaTeXStandardTheSubSection}%
\renewcommand{\thesubsubsection}{\LaTeXStandardTheSubSubSection}%
\renewcommand{\theparagraph}{\LaTeXStandardTheParagraph}%
}{}{}%

\@addtoreset{section}{titlecounter}

\usepackage{calc,amsfonts,amssymb,amsmath,bm,url,color,theorem,graphicx,cite}
\usepackage{algorithm}
\usepackage{algorithmic}
\usepackage{soul}
\usepackage{enumerate}
\usepackage{bbm}
\usepackage{shortcuts_OPT}
\usepackage{multirow}
\usepackage{subcaption}
\usepackage{array}

\usepackage{pifont}%

\usepackage{dblfloatfix}
\usepackage{todonotes}
\ifconfver \else \usepackage{titling} \fi

\usepackage{eqparbox}
\renewcommand\algorithmiccomment[1]{%
  ~\hfill$\triangleright$~\eqparbox{COMMENT}{#1}%
}

\definecolor{orange}{RGB}{255,107,0}

\newtheorem{Fact}{Fact}
\newtheorem{Lemma}{Lemma}

\newtheorem{Theorem}{Theorem}

\theorembodyfont{\rmfamily}

\newcommand\bw{\ensuremath{{\bm w}}}

\newcommand\bx{\ensuremath{{\bm x}}}
\newcommand\by{\ensuremath{{\bm y}}}

\newcommand\bh{\ensuremath{{\bm h}}}
\newcommand\bH{\ensuremath{{\bm H}}}

\newcommand\bz{\ensuremath{{\bm z}}}

\newcommand\bX{\ensuremath{{\bm X}}}

\newcommand\ba{\ensuremath{{\bm a}}}

\newcommand\bA{\ensuremath{{\bm A}}}

\newcommand\bb{\ensuremath{{\bm b}}}

\newcommand\bB{\ensuremath{{\bm B}}}

\newcommand\bd{\ensuremath{{\bm d}}}

\newcommand\bv{\ensuremath{{\bm v}}}

\newcommand\bs{\ensuremath{{\bm s}}}

\newcommand{\Rbb}{\mathbb{R}}
\newcommand{\Cbb}{\mathbb{C}}

\newcommand{\setS}{\mathcal{S}}
\newcommand{\setC}{\mathcal{C}}

\newcommand{\Exp}{\mathbb{E}}
\newcommand{\jj}{\mathfrak{j}}

\newcommand\br{\ensuremath{{\bm r}}}

\newcommand{\bI}{{\bm I}}

\newcolumntype{M}[1]{>{\centering\arraybackslash}m{#1}}

\usepackage{tikz}
\usetikzlibrary{arrows.meta}
\usetikzlibrary{decorations}
\usetikzlibrary{calc}
\usetikzlibrary{shapes.geometric}
\usetikzlibrary{external}

\DeclareMathSymbol{\mh}{\mathord}{operators}{`\-}

\begin{document}

\bibliographystyle{IEEEtran}

\newcommand{\papertitle}{
One-Bit MIMO  Detection: From Global Maximum-Likelihood Detector to Amplitude Retrieval Approach
}

\newcommand{\paperabstract}{
As communication systems advance towards the future 6G era, the incorporation of large-scale antenna arrays in base stations (BSs) presents challenges such as increased hardware costs and energy consumption.
To address these issues, the use of one-bit analog-to-digital converters (ADCs)/digital-to-analog converters (DACs) has gained significant attentions.
This paper focuses on one-bit multiple-input multiple-output (MIMO) detection in an uplink multiuser transmission scenario where the BS employs one-bit ADCs.
One-bit quantization retains only the sign information and loses the amplitude information, which poses a unique challenge in the corresponding detection problem.
The maximum-likelihood (ML) formulation of one-bit MIMO detection has a challenging likelihood function that hinders the application of many high-performance detectors developed for classic MIMO detection (under high-resolution ADCs).
While many approximate methods for the ML detection problem have been studied, it lacks an efficient \emph{global} algorithm.
This paper fills this gap by proposing an efficient branch-and-bound algorithm, which is guaranteed to find the global solution of the one-bit ML MIMO detection problem.
Additionally, a new amplitude retrieval (AR) detection approach is developed, incorporating explicit amplitude variables into the problem formulation.
 The AR approach yields simpler objective functions that enable the development of efficient algorithms offering both global and approximate solutions.
The paper also contributes to the computational complexity analysis of both ML and AR detection problems.
Extensive simulations are conducted to demonstrate the effectiveness and efficiency of the proposed formulations and algorithms.
}


\ifplainver


    \title{\papertitle}

    \author{Mingjie Shao, Wei-Kun Chen, Cheng-Yang Yu, Ya-Feng Liu, and Wing-Kin Ma}

    \maketitle

    \begin{abstract}
    \paperabstract
   \end{abstract}

\else
    \title{\papertitle}

    \ifconfver \else {\linespread{1.1} \rm \fi

    \author{Mingjie Shao, Wei-Kun Chen, Cheng-Yang Yu, Ya-Feng Liu, and Wing-Kin Ma
    }

    \maketitle

    \ifconfver \else
        \begin{center} \vspace*{-2\baselineskip}
        \end{center}
    \fi

    \begin{abstract}
    \paperabstract
    \end{abstract}


    \begin{IEEEkeywords}\vspace{-0.0cm}
 Amplitude retrieval,  global algorithm, maximum-likelihood, one-bit MIMO detection
    \end{IEEEkeywords}

    \ifconfver \else \IEEEpeerreviewmaketitle} \fi

 \fi

\ifconfver \else
    \ifplainver \else
        \newpage
\fi \fi

\makeatletter{\renewcommand*{\@makefnmark}{}
	\footnotetext{ Part of this work was presented at the IEEE SPAWC, 2023 \cite{yu2023efficient}. This work was completed during a visit of M. Shao and W.-K. Chen to the  State Key Laboratory of
Scientific and Engineering Computing (LSEC), Institute of Computational Mathematics and Scientific/Engineering Computing (ICMSEC), Academy of Mathematics and
Systems Science (AMSS), Chinese Academy of Sciences (CAS).
The work by M. Shao was supported by the Natural Science
Foundation of Shandong Province under Grant ZR2023QF103.
 The work of W.-K. Chen was supported by the National Natural Science Foundation of China (NSFC) under Grant 12101048.
  The work of Y.-F. Liu was supported in part by the NSFC under Grant 12371314 and Grant 12288201.
 The work by W.-K. Ma was supported by a General Research Fund of Hong Kong Research Grant Council under Project ID CUHK 14205421.}\makeatother}

\makeatletter{\renewcommand*{\@makefnmark}{}
	\footnotetext{ M. Shao is with the School of Information Science and
Engineering, Shandong University, Qingdao 266237, China (e-mail: mingjieshao@sdu.edu.cn).}\makeatother}

\makeatletter{\renewcommand*{\@makefnmark}{}
	\footnotetext{ W.-K. Chen and C.-Y. Yu are with the School of Mathematics and Statistics/Beijing Key
Laboratory on MCAACI, Beijing Institute of Technology, Beijing 100081,
China (e-mail: {chenweikun,~yuchengyang}@bit.edu.cn).}\makeatother}

\makeatletter{\renewcommand*{\@makefnmark}{}
	\footnotetext{ Y.-F. Liu is with the  LSEC, ICMSEC, AMSS, CAS, Beijing 100190, China (e-mail: yafliu@lsec.cc.ac.cn).}\makeatother}

\makeatletter{\renewcommand*{\@makefnmark}{}
	\footnotetext{ W.-K. Ma is with the Department of Electronic
Engineering, The Chinese University of Hong Kong, Hong Kong, China
(e-mail: wkma@ieee.org).}\makeatother}

\section{Introduction}
With the advancement of communication systems from 5G to future 6G, the base stations (BSs) are incorporating an increasing number of antennas to fulfill the demanding requirements of high spectrum efficiency and enhanced robustness.
However, the deployment of large-scale antenna arrays introduces challenges such as a higher number of analog-to-digital converters (ADCs)/digital-to-analog converters (DACs), and radio-frequency (RF) front ends. This, in turn, leads to elevated hardware costs and increased energy consumption at the BS.
To address these challenges, one promising approach is to replace high-resolution ADCs/DACs with low-resolution alternatives, especially one-bit ADCs/DACs, at the BS.
One-bit ADCs/DACs offer benefits such as cost-effectiveness and power efficiency.
Additionally, their output signals possess a constant envelope, which  is friendly in the sense of allowing better energy efficiency of the accompanying RF chains.

In light of the aforementioned background, coarse quantized signal processing has emerged as a topic of extensive investigation.
In conventional multiple-input multiple-output (MIMO) systems with  high-resolution ADCs/DACs, the quantization noise is usually small and hardly affects the system performance.
As a result, the quantization noise is typically considered negligible.
However, in MIMO systems with low-resolution ADCs/DACs, the quantization noise can be significant, and its effect should be better taken care of.
In the context of downlink communication with one-bit DACs, a prominent objective is to devise a methodology for designing coarsely quantized (discrete) transmitted signals that can satisfy some quality-of-service (QoS) requirements of multiple users.
This task, commonly referred to as one-bit precoding in the existing literature, typically manifests as a discrete optimization problem, prompting numerous research endeavors for algorithmic  designs and performance analysis;
see \cite{Jacobsson2017,Jedda2018,Sohrabi2018,shao2019framework,li2020interference,wu2023asymptotic} and the references therein.

In the multiuser uplink scenario employing one-bit ADCs, an important task pertains to MIMO detection in the presence of coarsely quantized observations, which is the focus of this study.
 The received signals at the BS only retain the sign information while losing the amplitude information.
Consequently, one-bit MIMO detection entails an inverse problem that involves detecting the multiuser signals from the one-bit quantized received signal.
Due to the strong nonlinearity and the amplitude loss associated with one-bit quantization, conventional MIMO detection methods developed for high-resolution ADCs cannot be directly applied to one-bit MIMO detection, or their performance may degrade~\cite{risi2014massive,choi2016near}.

The goal of this paper is to propose new formulations and efficient algorithms for the one-bit MIMO detection problem, which either is guaranteed to find the global solution of the considered problem or is able to strike a good balance between the detection performance and the computational cost.

\subsection{Related Works}
In view of the difficulties in one-bit MIMO detection, researchers have been exploring new methods to solve the problem.
Among the existing approaches, maximum-likelihood (ML) detection is perhaps the most widely studied
one. It is challenging to (globally) solve the ML detection problem because the problem is a nonlinear integer programming (NLIP) problem and the likelihood function involves integrals that do not admit an explicit form.
To tackle the ML problem efficiently, several approximate techniques have been proposed, including proximal gradient algorithms \cite{choi2016near,studer2016quantized,shao2020binary,Mirfarshbafan2020}, expectation maximization (EM) methods \cite{plabst2018efficient,shao2024accelerated}, search-based algorithms~\cite{jeon2018one}, and deep learning methods \cite{shao2020binary,nguyen2021linear,khobahi2021lord,shao2023explanation}.
Nonetheless, it should be noted that simple convex relaxation~\cite{choi2016near,Mirfarshbafan2020} or Gaussian approximation~\cite{plabst2018efficient} of the constellation symbols can significantly degrade the detection performance.
Therefore, researchers have explored various strategies to narrow the performance gap with the ML solution, such as employing extreme point pursuit~\cite{shao2020binary}, refining the solution through local search techniques~\cite{choi2016near,nguyen2021linear}, and leveraging the potential of deep learning~\cite{shao2020binary,nguyen2021linear,khobahi2021lord,shao2024accelerated}.
In addition, there are works that try to approximate the thorny likelihood function.
For instance, the works \cite{jeon2018one} and \cite{yi20231} apply local quadratic approximations, while \cite{nguyen2021linear} considers to use the sigmoid function to approximate the likelihood function in order to simplify the deep network training.
Despite these efforts, the development of an efficient {\it globally optimal} method for the one-bit  ML MIMO detection problem remains a challenge; in particular,  the highly nonlinear likelihood function  stands as an obstacle in the design of globally optimal algorithms.
Consequently, neither can classic MIMO detection techniques relying heavily on the quadratic objective form, such as the sphere decoding algorithm \cite{damen2003maximum}, nor off-the-shelf optimization solvers like CPLEX \cite{CPLEX2022} be readily exploited in this context.

In addition to ML detection, researchers have also explored alternative possibilities for MIMO detection under one-bit quantization.
Linear detectors, known for their simplicity, have been extensively studied.
The focus of this line of research lies in analyzing the impact of quantization on linear detectors such as maximum ratio combination (MRC) and zero-forcing (ZF) \cite{choi2015quantized,li2017channel}.
Moreover, modified linear receivers leveraging the Bussgang theorem have been proposed with the aim of improving the detection performance \cite{nguyen2021linear}.
Additionally, the one-bit MIMO detection can be interpreted as a binary classification task, which has been handled by a support vector machine formulation \cite{nguyen2021svm}.
Another closely related approach to ML detection is maximum a posteriori (MAP) detection.
Researchers have investigated various methods to approximate MAP detection, including the approximate message passing technique \cite{wen2015bayes} and the variational inference method~\cite{thoota2021variational}.
These approaches share similarities with EM methods used in ML detection, in terms of both their key steps and computational complexities.
In addition, spatial sigma-delta modulation applies the principles of antenna feedback and noise shaping to mitigate the quantization noise, which provides a different way for receiver design~\cite{Hessam2020Spectral}.

\subsection{Our Contributions}

In this paper, we study one-bit MIMO detection in a single-cell multiuser uplink scenario.
We start with the development of a global algorithm, which is an efficient branch-and-bound algorithm, for solving the ML detection problem.
In addition, in order to bypass   sophisticated optimization techniques to handle the highly nonlinear likelihood function of the ML detection problem, we propose to incorporate the missing amplitude information into the one-bit MIMO detection problem.
This leads to a new one-bit MIMO detection formulation that is much easier to handle than the one-bit ML MIMO detection formulation.
We provide extensive simulations to demonstrate the detection performance and computational complexities of the proposed approaches.

The contributions of this work are summarized as follows.
\begin{enumerate}
  \item [(i)] \emph{First Global Algorithm for One-Bit ML MIMO Detection}.
  We  first analyze the complexity status of the one-bit ML MIMO detection problem.
    It is well-known that the classic ML MIMO detection problem is NP-hard \cite{verdu1989computational}, but the complexity status of the one-bit ML MIMO detection problem remains unknown.
    We fill this theoretical gap by showing that the one-bit ML MIMO detection problem is NP-hard.
    Then, we propose the \emph{first} global algorithm for one-bit ML MIMO detection. Specifically, we first transform the one-bit ML MIMO detection problem into an equivalent mixed integer linear programming (MILP) problem with an exponential number of constraints (with respect to the number of users).
  To solve the proposed MILP problem, we employ a delayed constraint generation framework, which starts with a relaxed MILP problem with only a selected small subset of  constraints, and gradually adds the neglected constraints (when needed) until an optimal solution of the original problem is found.
  In order to develop a lightweight global algorithm, we solve each relaxed MILP problem inexactly, which is achieved by embedding the delayed constraint generation procedure into the branch-and-bound procedure.
  In this way, the algorithm only needs to solve linear programming (LP) subproblems with significantly  smaller problem sizes (compared with the MILP reformulation of the original problem), and thus is computationally efficient.


\item[(ii)] \emph{New AR Formulation and Low-Complexity Algorithm}.
 To address the loss of the amplitude information due to quantization, we propose an alternative formulation to one-bit ML MIMO detection by introducing an explicit amplitude variable into the problem.
 The key feature of the amplitude retrieval (AR) formulation is that it has much simpler objective functions (quadratic or linear) compared to ML detection, which significantly facilitates the design of  computationally efficient  global and approximate algorithms.
    In particular, we can directly apply the state-of-the-art optimization solvers to the AR formulation to find an optimal solution.
    In addition, we leverage the alternate Barzilai-Borwein (ABB) method \cite{dai2005projected}, which is a first-order projected gradient (PG) method with a modified step size, to obtain a computationally efficient solution.


\end{enumerate}

We provide extensive simulations to test the efficacy of the proposed algorithms.
The simulation results show that the proposed global algorithm for the one-bit ML MIMO detection problem can obtain an optimal solution by a considerably reduced runtime than the exhaustive search, which enables it to be an important performance benchmark for various existing approximate algorithms developed for the same problem.
Moreover, we provide numerical evidence to show that the AR formulation is a reasonable alternative to the ML formulation.
Compared to the ML formulation, the AR formulation, which is tackled by our custom-built algorithm,  can yield competitive   bit-error rate (BER) performance with a much lower computational complexity.

Part of this  paper was presented in a conference \cite{yu2023efficient}. It studied the global algorithm design for one-bit ML MIMO detection, which corresponds to Subsections \ref{sec:eff_alg} and \ref{sec:BnB} in this paper.
Compared to its conference version \cite{yu2023efficient}, this paper has many new contributions in  complexity  analysis,  problem formulation, and algorithmic design.
First, this  paper presents the complexity analysis for one-bit ML MIMO detection, which was not considered in \cite{yu2023efficient}.
More importantly, this paper develops new AR formulations and efficient algorithms, which stands as a new core contribution of this paper.

\subsection{Organization and Notations}

Our paper is organized as follows.
Section~\ref{sect:em_basics} reviews the one-bit ML MIMO detection problem.
Section~\ref{sec:ML_alg} studies the complexity analysis of one-bit ML MIMO detection problem, and proposes a global algorithm for  solving it.
Section~\ref{sec:AR} describes the AR formulation, analyzes its complexity, and presents a custom-built algorithm for solving it.
Section~\ref{sec:sim} shows   extensive simulation results to illustrate the performance of the proposed algorithms.
Section~\ref{sec:conc} draws the conclusion.

Our adopted notations are standard.
We use $\Rbb$ and $\Cbb$ to denote the real and complex space, respectively.
The boldface lowercase letters, e.g., $\bx$, represent vectors; $x_i$ represents the $i$th element of $\bx$;
the boldface uppercase letters, e.g., $\bX$, represent matrix;
$|\setS|$ denotes the cardinality of a set $\setS$;
$\Re(x)$ and $\Im(x)$ denote the real and imaginary parts of $x$, respectively;
$\langle \bx, \by \rangle$ denotes the inner product between $\bx$ and $\by$;
$\bX^{\top}$ and  $\bX^{\dag}$  denote the transpose and  pseudo-inverse, respectively;
$\mbox{Diag}(\bx)$ represents a diagonal matrix with $\bx$ being the diagonal elements;
$\| \bx \|_{n}$ denotes the $\ell_n$-norm of $\bx$ for $n\geq 1$;
${\cal N}(\bm \mu, \bm C)$ represents the Gaussian distribution with mean $\bm \mu$ and covariance $\bm C$.

\section{One-Bit ML MIMO Detection}
\label{sect:em_basics}

This section presents the probit signal model and one-bit ML MIMO detection problem, which paves the way to the global algorithmic design in Section \ref{sec:ML_alg}.

\subsection{Signal Model}

The problem of interest can be posed as the following probit model
\begin{equation}\label{eq:model}
  \begin{split}
    \br = \mbox{sgn}(\by), \quad \by = \bH \bx + \bv.
  \end{split}
\end{equation}
Here, $\br\in \Rbb^M$ is the observation vector;
the function $\mbox{sgn}$ takes the sign of its argument, thus the elements of $\br$ are  binary, i.e., $r_i \in \{ -1,1 \}$ for $i=1,2,\ldots,M$;
$\by\in \Rbb^M$ is the unquantized counterpart of $\br$, but is not observable;
$\bH\in \Rbb^{M\times N}$ is a system matrix;
$\bx\in\Rbb^{N}$ is an unknown binary variable, i.e., $x_i \in\{ -1,1 \}$ for  $i =1,2,\ldots, N$;
and $\bv\in \Rbb^{M}$ is a white Gaussian noise vector with $\bv \sim {\cal N}(\bm 0, \sigma^2 \bI)$.
The problem  is to infer the  variable $\bx$ from the binary observation $\br$, given the information of $\bH$ and $\sigma$.

Let us delineate how the interested one-bit MIMO detection problem falls into the above probit model.
Consider a massive MIMO system where the BS employs a pair of one-bit ADCs at each antenna out of the consideration of hardware cost and power consumption.
In the uplink transmission,  a number of $\tilde{N}$ single-antenna users concurrently send their signals to the BS with $\tilde{M}$ antennas.
The received signal can be modeled by
\begin{equation}\label{eq:model_com}
\begin{split}
  \tilde{\br} =  ~ {\cal Q} (\tilde{\by}),~~\tilde{\by} =  ~ \tilde{\bH} \tilde{\bx} +\tilde{\bv},
 \end{split}
\end{equation}
where $\tilde{\bx}\in \Cbb^{\tilde{N}}$ is the multiuser transmit signal vector, whose elements are assumed to be drawn from the Quadrature Phase Shift Keying (QPSK) constellation $\{ \pm 1 \pm \jj \}$;
$\tilde{\bH} \in \Cbb^{\tilde{M} \times \tilde{N}}$ is the multiuser channel matrix;
$\tilde{\bv}\in \Cbb^{\tilde{M}}$ is   additive complex Gaussian noise with mean $\bm 0$ and covariance matrix $\tilde{\sigma}^2 \bI$;
\[
 {\cal Q}(x): = \mbox{sgn}(\Re(x))+ \jj\cdot \mbox{sgn}(\Im(x))
\]
is the one-bit quantizer associated with the one-bit ADCs for both the real and imaginary parts of $x$;
 $ \tilde{\br}\in \Cbb^{\tilde{M}}$ is the received one-bit signal at the BS.
One-bit MIMO detection aims to detect the multiuser signal $\tilde{\bx}$ from the received  one-bit signal $\tilde{\br}$, given the information of $\tilde{\bH}$ and $\tilde{\sigma}^2$.

Define
\begin{equation*}
  \begin{split}
    \by  =    \begin{bmatrix}
            \Re(\tilde{\by})\\
             \Im(\tilde{\by})
           \end{bmatrix} \in \Rbb^M,  \bH   =    \begin{bmatrix}
                                   \Re(\tilde{\bH})  &     -\Im(\tilde{\bH}) \\
                                  \Im(\tilde{\bH}) &   \Re(\tilde{\bH})
                                 \end{bmatrix} \in \Rbb^{M\times N},\\
    \bx   =   \begin{bmatrix}
            \Re(\tilde{\bx})\\
             \Im(\tilde{\bx})
           \end{bmatrix}\in \Rbb^N,
            \br   =   \begin{bmatrix}
            \Re(\tilde{\br})\\
             \Im(\tilde{\br})
           \end{bmatrix} \in \Rbb^M,
                 \bv   =   \begin{bmatrix}
            \Re(\tilde{\bv})\\
             \Im(\tilde{\bv})
           \end{bmatrix} \in \Rbb^M,
  \end{split}
\end{equation*}
where $M=2\tilde{M}$, $N = 2\tilde{N}$, and $\bv\in\Rbb^{M}$ follows the standard Gaussian distribution with mean $\bm 0$ and covariance $\sigma^2 \bI$ with $\sigma^2 = \tilde{\sigma}^2/2$.
With this transformation,  the one-bit MIMO detection problem is a special case of the probit model in~\eqref{eq:model}.

\subsection{ ML Detection}

We consider the ML detection associated with  the probit model \eqref{eq:model}.
The ML detection problem can be formulated as~\cite{choi2016near}
\begin{equation}\label{eq:ML}
  \min_{\bx\in \{ -1,1 \}^N} f(\bx): =-\sum_{i=1}^{M} \log  \Phi \left(\frac{\bb_i^{\top}\bx}{\sigma} \right),
\end{equation}
where $\bb_i^\top = r_i \cdot \bh_i^{\top}$, $\bh_i^{\top}$ denotes the $i$th row of $\bH$, {$\Phi(z) = \int_{-\infty}^{z} \frac{1}{\sqrt{2\pi}}e^{-t^2}\ dt$} is the cumulative distribution function of the standard Gaussian distribution, and $f(\bx)$ is the negative log-likelihood function.

The one-bit ML MIMO detection problem \eqref{eq:ML} is an NLIP problem.
The difficulty of solving problem \eqref{eq:ML} arises from two aspects: (i) the objective function $f(\bx)$ involves integrals  that do not have   closed-form expressions; (ii) the decision variables $\bx$ are binary, whose dimension could be large.
One could apply an exhaustive search to globally solve \eqref{eq:ML}, which needs to examine all feasible solutions with a complexity order of ${\cal O}(2^N)$  ---which is exponentially increasing with the number of users.
The computational complexity can be unaffordable in a massive MIMO system where the number of users can be tens or more.
Moreover, to the best of our knowledge,  off-the-shelf efficient mixed integer linear programming (MILP) solvers such as CPLEX \cite{CPLEX2022} cannot directly handle the integral $\Phi(\cdot)$.

One may wonder whether the rich results in classic MIMO detection \cite{tan2001application,damen2003maximum,ma2002quasi,wubben2011lattice,lu2019tightness,yang2015fifty,zhao2021efficient}, where high-resolution ADCs are employed at the BS and $\by$ is directly accessible, can be applied to the one-bit ML MIMO detection problem.
Unfortunately, the classic ML MIMO detection problem has a different  objective function form (it is quadratic), and many high-performance classic MIMO detectors, including lattice reduction \cite{wubben2011lattice}, sphere decoding \cite{damen2003maximum} and semidefinite relaxation \cite{tan2001application,ma2002quasi}, were developed based on the latter.
In the literature, researchers have proposed many approximate algorithms for solving the one-bit ML MIMO detection problem \eqref{eq:ML}
that seek to strike a balance between the detection performance and the computational complexity \cite{choi2015quantized,choi2016near,shao2020binary,Mirfarshbafan2020,nguyen2021linear,khobahi2021lord ,plabst2018efficient,plabst2018efficient,shao2024accelerated}.
Many of them apply convex relaxation methods on the binary constraints to avoid solving a problem with integer variables~\cite{choi2016near,shao2020binary,Mirfarshbafan2020,plabst2018efficient}.
Unfortunately, these algorithms do not have a guarantee to  retrieve the globally optimal solution to the one-bit ML MIMO detection problem~\eqref{eq:ML}.

In this paper, we aim to overcome the above weaknesses by
proposing an efficient algorithm for globally solving problem~\eqref{eq:ML} in Section \ref{sec:ML_alg}. Moreover, we will explore new yet simple alternative   formulations for one-bit MIMO detection in Section~\ref{sec:AR}.

\section{An Efficient Global Algorithm}
\label{sec:ML_alg}

This section presents the complexity analysis of the one-bit ML MIMO detection problem \eqref{eq:ML}   and an efficient dedicated global algorithm for solving it.

\subsection{Complexity Analysis}

We first analyze the complexity
status of problem \eqref{eq:ML}.
It is well-known that the classic ML MIMO detection problem is NP-hard \cite{verdu1989computational}.
However, it is still unknown whether there exists a polynomial time algorithm for solving problem \eqref{eq:ML}.
We fill this theoretical gap by showing the following theorem.
\begin{Theorem}\label{thm:ML_NP}
 	The one-bit ML MIMO detection problem \eqref{eq:ML} is NP-hard.
\end{Theorem}
To prove Theorem \ref{thm:ML_NP}, we need the following lemma.
\begin{Lemma}\label{Lem:ML_NP}
    Let $\Phi(x) = \frac{1}{\sqrt{2\pi}} \int_{-\infty}^{x} e^{-\frac{t^2}{2}}\ dt$.
    Then, it holds that
    \begin{equation}
	\min_{x \in \mathbb{R}} -\log\Phi (x) - \log\Phi(-x) = -2\log\Phi(0),
    \end{equation}
    where the only minimum point is arrived at $x=0$.
\end{Lemma}
\noindent{\it Proof of Lemma~\ref{Lem:ML_NP}}:
	Observe that
\begin{equation*}
	\begin{aligned}
        & ~\Phi (x)\cdot \Phi(-x)\\
	 = &\left(\frac{1}{\sqrt{2\pi}} \int_{-\infty}^{x} e^{-\frac{t^2}{2}}\ dt\right) \left(\frac{1}{\sqrt{2\pi}} \int_{-\infty}^{-x} e^{-\frac{t^2}{2}}\ dt\right) \\
	  =& \left[\Phi(0)+ \frac{1}{\sqrt{2\pi}} \int_{0}^{x} e^{-\frac{t^2}{2}}\ dt\right] \left[\Phi(0)+ \frac{1}{\sqrt{2\pi}} \int_{0}^{-x} e^{-\frac{t^2}{2}}\ dt\right] \\
	   =& \left[\Phi(0)+ \frac{1}{\sqrt{2\pi}} \int_{0}^{x} e^{-\frac{t^2}{2}}\ dt\right] \left[\Phi(0)- \frac{1}{\sqrt{2\pi}} \int_{0}^{x} e^{-\frac{t^2}{2}}\ dt\right] \\
	=& {[\Phi(0)]}^2 -{ \left(\frac{1}{\sqrt{2\pi}} \int_{0}^{x} e^{-\frac{t^2}{2}}\ dt\right)}^2   \\
     \leq & {[\Phi(0)]}^2,
	\end{aligned}
\end{equation*}
where {the inequality holds with equality if and only if $x=0$}.
This, together with the fact that $-\log\Phi (x) - \log\Phi(-x) = - \log[\Phi (x) \cdot \Phi(-x)]$, shows the desired result. \hfill $\blacksquare$

\medskip

\noindent {\it Proof of Theorem \ref{thm:ML_NP}}.
We  prove Theorem \ref{thm:ML_NP} by showing   that there exists a special instance of problem \eqref{eq:ML} which is as hard as  an NP-complete partition problem \cite{Garey1978}:
given a finite set $\mathcal{S}=\{1,2,\ldots, n\}$ and a size $w_i \in \mathbb{Z}_+$ for the $i$-th element with $\sum_{i \in\mathcal{S}} w_i = 2W$, does there exist a partition $\mathcal{S}=\mathcal{S}_1 \cup  \mathcal{S}_2$ with $\setS_1 \cap \setS_2 = \emptyset$ such that $\sum_{i \in \mathcal{S}_1} w_i = \sum_{i \in \mathcal{S}_2} w_i = W$?
	
Given any instance of the partition problem, we construct an instance of problem \eqref{eq:ML} by setting $M=2$, $N=n$, $\sigma =1$, $\bb_1 = \bw$, and $\bb_2 = -\bw$.
By construction, problem \eqref{eq:ML} reduces to
\begin{equation}\label{eq:MLex}
   { v^{\ast}}= \min_{\bx \in \{ -1,1 \}^n} -\log\Phi (\bw^\top\bx) - \log\Phi( -\bw^\top\bx).
\end{equation}
From Lemma \ref{Lem:ML_NP}, $ { v^{\ast}}=  -2\log\Phi(0)$ if and only if $\bw^\top \bx=0$ holds for some $\bx \in \{-1,1\}^n $, which is further equivalent to the existence of the partition  $\mathcal{S}=\mathcal{S}_1 \cup  \mathcal{S}_2$ with $\sum_{i \in \mathcal{S}_1} w_i = \sum_{i \in \mathcal{S}_2} w_i = W$, that is, the answer to the partition problem is yes.
The above transformation can be done in polynomial time.
Since the partition problem is NP-complete, we conclude that problem \eqref{eq:ML} is NP-hard.  \hfill $\blacksquare$
\medskip

Theorem \ref{thm:ML_NP} reveals the intrinsic difficulty of (globally) solving problem \eqref{eq:ML}. In particular,  there does not exist a polynomial time algorithm for solving problem \eqref{eq:ML} unless $\text{P}=\text{NP}$.
In the next subsection, we will design an efficient algorithm for globally solving problem \eqref{eq:ML}.


\subsection{ An Efficient Global Algorithm}
\label{sec:eff_alg}


In this subsection, we present a global algorithm for solving problem \eqref{eq:ML}.
To do this, we first equivalently reformulate the NLIP  problem \eqref{eq:ML} as an MILP problem with an exponential number of constraints.
Then we apply the delayed constraint generation procedure \cite{Bertsimas1997} to solve the formulated MILP problem in which only a small subset of constraints is initially considered, and additional constraints are gradually added (when needed) until an optimal solution of the original problem is found.
Finally, to speed up the solution process, we integrate the delayed constraint generation procedure into the branch-and-bound algorithm, resulting into a customized efficient global algorithm for solving problem \eqref{eq:ML}.

\smallskip

\subsubsection{An MILP  Reformulation}
We first equivalently reformulate problem \eqref{eq:ML} as
\begin{equation}\label{eq:ML2}
\begin{split}
  \min_{\bx, \bw} &~ \sum_{i=1}^{M} w_i \\
  \mbox{s.t. } &~  w_i\geq  g_i(\bx),  ~\forall~i=1,2,\ldots, M ,\\
  &~ \bx\in \{ -1,1 \}^N,
  \end{split}
\end{equation}
where $\bw = {[ w_1,w_2,\ldots, w_M ]^{\top}}$ and
\[
    g_i(\bx): = -\log  \Phi \left(\frac{\bb_i^{\top}\bx}{\sigma} \right).
\]
Note that $g_i$ is a convex function with respect to $\bx$.
 Thus, by Jensen's inequality, we obtain the following linear inequality
\begin{equation}\label{eq:convex}
    g_i(\bx) \geq g_i(\hat{\bx}) + \langle \nabla g_i(\hat{\bx}), \bx - \hat{\bx}\rangle,~ \forall ~ \hat{\bx} \in \{ -1,1 \}^N,
\end{equation}
where
\[
    \nabla g_i(\bx) = -\frac{\phi(\bb_i^{\top}\bx/\sigma)}{\Phi(\bb_i^{\top}\bx/\sigma)}\frac{\bb_i}{\sigma}
\]
is the gradient of $g_i$ at $\bx$, {and} $\phi(t) = \frac{1}{\sqrt{2\pi}}e^{-t^2/2}$ is the probability distribution function of the standard Gaussian distribution.
Then, with \eqref{eq:convex}, we can reformulate problem \eqref{eq:ML2} as follows
\begin{subequations}\label{eq:ML3}
\begin{align}
  (\bx^{\star}, \bw^{\star}) =\arg \min_{\bx, \bw} &~ \sum_{i=1}^{M} w_i \notag\\
  \mbox{s.t. } &~  w_i\geq g_i(\hat{\bx}) + \langle\nabla g_i(\hat{\bx}),\bx - \hat{\bx}\rangle, \label{eq:lin_ineq}\\
  &~~  \forall~ i=1,2,\ldots, M, ~\forall~\hat{\bx} \in \{ -1,1 \}^N, \notag\\
  &~ \bx\in \{ -1,1 \}^N.
  \end{align}
\end{subequations}
\begin{Fact}\label{fact:eqv}
  Problems \eqref{eq:ML} and \eqref{eq:ML3} are equivalent, in the sense that they have the same optimal solution for $\bx$.
\end{Fact}
Fact~\ref{fact:eqv} can be obtained by noting that inequality \eqref{eq:convex} is tight when $\hat{\bx} = \bx$, which establishes the equivalence between problems \eqref{eq:ML2} and \eqref{eq:ML3}. This, together with the equivalence between \eqref{eq:ML} and \eqref{eq:ML2},   leads to the desired result.

The upshot of problem \eqref{eq:ML3} is that the inequalities \eqref{eq:lin_ineq} are \emph{linear} in both $\bx$ and $\bw$.
As a result, problem \eqref{eq:ML3} is an MILP problem.
In principle, problem \eqref{eq:ML3}  can be solved by off-the-shelf MILP solvers such as CPLEX \cite{CPLEX2022}.
However, the total number of inequality constraints in \eqref{eq:lin_ineq}  is $M\cdot 2^N$, where both $M$ and $N$ can be large in massive MIMO systems, which can lead to prohibitively high computational complexity.

\smallskip
\subsubsection{A Delayed Constraint  Generation Framework}

To address the computational challenge arising from the exponential number of constraints in \eqref{eq:lin_ineq}, we employ the delayed constraint generation framework.
This framework solves problem \eqref{eq:ML3} by initially considering a small subset of constraints in \eqref{eq:lin_ineq} and gradually adding the neglected constraints (when needed) until an optimal solution of the problem is found.
For more details of the delayed constraint generation framework, we refer to \cite{Bertsimas1997}.
In the following, we detail the delayed constraint generation framework to solve problem \eqref{eq:ML3}.


Define the index set
\[
\setC = \{ (i,\hat{\bx})~| ~i=1,2,\ldots, M, ~  \hat{\bx} \in \{ -1,1 \}^N \}
\]
and select $\setS \subseteq \setC$ as a subset of $\setC$.
We consider the following relaxation of problem \eqref{eq:ML3}:
\begin{equation}\label{eq:rel}
\begin{split}
\min_{\bx, \bw} &~\sum_{i=1}^{M} w_i \\
  \mbox{s.t. }&~ w_i\geq g_i(\hat{\bx}) + \langle \nabla g_i(\hat{\bx}),\bx - \hat{\bx}\rangle, ~\forall~(i,\hat{\bx}) \in \setS,\\
  &\quad \quad \bx\in \{ -1,1 \}^N.
  \end{split}
\end{equation}
Denote by $(\bar{\bx}, \bar{\bw}) $ the optimal solution to problem \eqref{eq:rel}.
We have the following result.
\begin{Fact}\label{fact:rex}
  Consider problems \eqref{eq:ML3} and \eqref{eq:rel}. The following statements hold.
  \begin{itemize}
  \item [(i)] $\sum_{i=1}^{M}\bar{w}_i\leq \sum_{i=1}^{M} w_i^{\star}$;
  \smallskip
\item [(ii)]  if $\bar{w}_i \geq g_{i}(\bar{\bx})$ holds for all $i$, then $(\bar{\bx}, \bar{\bw})$ is also optimal to problem \eqref{eq:ML3}.
    \end{itemize}
\end{Fact}
{\it Proof}:
Since problem \eqref{eq:rel} is a relaxed version of problem \eqref{eq:ML3}, it holds that $\sum_{i=1}^{M}\bar{w}_i\leq \sum_{i=1}^{M} w_i^{\star}$. This proves a).

If $\bar{w}_i \geq g_{i}(\bar{\bx})$ for all $i$, then $(\bar{\bx}, \bar{\bw})$ is a feasible solution to problem \eqref{eq:ML2}.
Hence we get $\sum_{i=1}^M \bar{w}_i \geq  \sum_{i=1}^M w^{\star}_i$.
This, together with a), implies $\sum_{i=1}^M \bar{w}_i =  \sum_{i=1}^M w^{\star}_i$.
Therefore, $(\bar{\bx}, \bar{\bw})$ is an optimal solution to problem \eqref{eq:ML2}, and also problem~\eqref{eq:ML3}. \hfill $\blacksquare$

\medskip

Fact~\ref{fact:rex} offers a hint to the algorithmic design.
Specifically, we start from solving problem \eqref{eq:rel} with an $\setS\subseteq \setC$.
If $\bar{w}_i \geq g_{i}(\bar{\bx})$ holds for all $i$, then $(\bar{\bx}, \bar{\bw})$ is already optimal to problem \eqref{eq:ML3}.
Otherwise, if $\bar{w}_i < g_{i}(\bar{\bx})$ for some $i$, the constraint
\begin{equation}\label{eq:cons}
 w_i\geq g_i(\bar{\bx}) + \langle\nabla g_i(\bar{\bx}),\bx - \bar{\bx}\rangle
\end{equation}
is added into problem \eqref{eq:rel}, i.e., adding $(i, \bar{\bx})$ into $\setS$.
Then, we solve problem \eqref{eq:rel} again with the updated $\setS$.
This process is repeated until $\bar{w}_i \geq g_{i}(\bar{\bx})$ holds for all $i$.
This delayed constraint generation framework is described in Algorithm~\ref{Alg:OTF}.


\begin{algorithm}[t!]
\caption{A Delayed Constraint Generation Framework for Solving Problem \eqref{eq:ML3}}\label{Alg:OTF}
\begin{algorithmic}[1]
\renewcommand{\algorithmiccomment}[1]{~~//\,\texttt{#1}}
\STATE {\bf input:} Initialization $\setS \subseteq \setC$;
\REPEAT[{one iteration}]
\STATE Solve problem \eqref{eq:rel} to obtain its optimal solution $(\bar{\bx}, \bar{\bw})$;

\IF {$\bar{w}_i < g_{i}(\bar{\bx})$  for some $i$'s}
\STATE $\setS \leftarrow \setS  \cup  \{(i, \bar{\bx}) \mid  \bar{w}_i < g_{i}(\bar{\bx}), ~ i=1,2,\ldots, M \}$;
\ELSE
\STATE {\bf break};

\ENDIF

\UNTIL {$\bar{w}_i \geq g_{i}(\bar{\bx})$ holds for all $i$};

\STATE {\bf output}  $(\bx^{\star}, \bw^{\star}) = (\bar{\bx}, \bar{\bw})$.

\end{algorithmic}
\end{algorithm}

%

\subsection{An Efficient Branch-and-Bound Algorithm}
\label{sec:BnB}

Algorithm~\ref{Alg:OTF} requires (possibly) solving multiple MILP problems in the form of \eqref{eq:rel}
(e.g., by branch-and-bound algorithms \cite{Achterberg2009}) and solving each MILP problem can be time-consuming. Therefore, the complexity of Algorithm 1 can still be high (especially when the number of iterations is large).
To further reduce the computational complexity of Algorithm~\ref{Alg:OTF}, we propose to solve each {MILP problem} \eqref{eq:rel} {inexactly}, which is done by embedding the delayed constraint generation procedure (cf. lines 3-8 in Algorithm~\ref{Alg:OTF}) into one branch-and-bound algorithm.
Branch-and-bound algorithms are tree search methods that recursively partition the feasible region (i.e., a rooted tree) into small subregions (i.e., branches).
In particular, our proposed branch-and-bound algorithm solves the  LP  relaxation in {the} form of \eqref{eq:rel2} at each iteration and gradually tightens the relaxation by adding appropriate $(i, {\hat{\bx}})$ in the set {$\setS$} and fixing more elements of {$\bx$} to be $\{-1, 1\}.$
The resulting algorithm is still a \emph{global} algorithm to problem \eqref{eq:ML3}.
Note that the proposed algorithm only needs to solve an LP problem at each iteration, which is in sharp contrast to solving {the MILP problem} \eqref{eq:rel} in Algorithm \ref{Alg:OTF}.
Below, we present the proposed algorithm in more details.

\smallskip
\subsubsection{Subproblems and Their LP Relaxations}

Denote $\mathcal{F}_+$ and $\mathcal{F}_-$ as some subsets of $\{1, 2, \ldots, N\}$ such that ${x}_j=1$ for $j \in \mathcal{F}_+$ and  ${x}_j=-1$ for $j \in \mathcal{F}_-$, and $\mathcal{F}_+ \cap \mathcal{F}_-=\varnothing$.
The subproblem to explore at the branch defined by $\mathcal{F}_+$ and $\mathcal{F}_-$ is given by
\begin{subequations}\label{eq:ML3r}
	\begin{align}
		\min_{\bx, \bw} &~ \sum_{i=1}^{M} w_i \notag\\
		\mbox{s.t. } &~  w_i\geq g_i(\hat{\bx}) + \langle\nabla g_i(\hat{\bx}),\bx - \hat{\bx}\rangle,  ~\forall~(i,\hat{\bx}) \in \setC,\label{eq:lin_ineq1}\\
		& ~x_j = 1, ~\forall~j \in \mathcal{F}_+, ~x_j =-1, ~\forall ~j \in \mathcal{F}_-,\\
		&~ \bx\in \{ -1,1 \}^N. \label{eq:bin}
	\end{align}
\end{subequations}
{Also, consider the following LP relaxation of problem \eqref{eq:ML3r}:}
\begin{subequations}\label{eq:rel2}
	\begin{align}
		\min_{\bx, \bw}& \sum_{i=1}^{M} w_i \notag\\
		\mbox{s.t. } &~  w_i\geq g_i(\hat{\bx}) + \langle\nabla g_i(\hat{\bx}),\bx - \hat{\bx}\rangle, ~\forall~(i,\hat{\bx}) \in \setS,\label{eq:lin_ineq2}\\
		& ~x_j = 1,  ~\forall~j \in \mathcal{F}_+, ~x_j =-1,  ~\forall~j \in \mathcal{F}_-,\\
		&~ \bx\in [-1,1 ]^N, \label{eq:cont}
	\end{align}
\end{subequations}
where $\setS \subseteq \setC$.
Problem \eqref{eq:rel2} is a relaxation of problem \eqref{eq:ML3r} by replacing $\setC$ with $\setS$ and by relaxing binary variables $x_j$'s with $j \notin\mathcal{F}_+\cup\mathcal{F}_-$ to $[-1,1]$.
Therefore, solving the LP problem \eqref{eq:rel2} provides a lower bound for the MILP problem \eqref{eq:ML3r}.

\smallskip
\subsubsection{Proposed Algorithm}
Now, we present the main steps of the proposed branch-and-bound algorithm based on the LP relaxation in \eqref{eq:rel2}.
We use $(\check{\bx},\check{\bw})$ to denote the best-known feasible solution that provides the smallest objective value at the current iteration and use $U$ to denote its objective value (called the \emph{upper bound} of problem \eqref{eq:ML3}).
In addition, {we use $(\mathcal{F}_+, \mathcal{F}_-, \setS)$ to denote subproblem \eqref{eq:ML3r} where $\setS \subseteq \setC$  corresponds to its current LP relaxation \eqref{eq:rel2}}, and
$\mathcal{P}$ to denote the problem set of the current unprocessed subproblems.
At the beginning, we initialize $\mathcal{P}\leftarrow\{(\varnothing, \varnothing, \setS)\}$  for some $\setS \subseteq \setC$.
At each iteration, we pick a subproblem $(\mathcal{F}_+, \mathcal{F}_-, \setS)$ from $\mathcal{P}$, and solve problem \eqref{eq:rel2} to obtain its solution $(\bx_{\sf LP}, \bw_{\sf LP})  $ and objective value $f_{\sf LP} = \sum_{i=1}^N [\bw_{\sf LP}]_i $.
Then, one of the following cases must happen:
\begin{itemize}
	\item [(i)] If $f_{\sf LP} \geq U$, then problem \eqref{eq:ML3r} cannot contain a feasible solution that provides an objective value better than $U$ (and this subproblem does not need to be explored).
	\item [(ii)] If $f_{\sf LP} < U$ and $\bx_{\sf LP} \in \{-1,1\}^N$, there are two subcases.
	\begin{itemize}
		\item [(ii.1)] If $[\bw_{\sf LP}]_i \geq g_i (\bx_{\sf LP} )$ for all $i=1,2,\ldots,M$,  then $(\bx_{\sf LP}, \bw_{\sf LP})$ must be an optimal solution to problem \eqref{eq:ML3r}.
We update $ (\check{\bx},\check{\bw}) \leftarrow (\bx_{\sf LP}, \bw_{\sf LP})$ and $U \leftarrow f_{\sf LP}$.
		\item [(ii.2)] Otherwise, we apply the delayed constraint generation procedure by adding $(i, \bx_{\sf LP})$ with all $ [\bw_{\sf LP}]_i < g_{i}(\bx_{\sf LP})$ into $\setS$ to obtain a tightened problem \eqref{eq:rel2}.
	\end{itemize}
	\item [(iii)] If $f_{\sf LP} < U$ and $\bx_{\sf LP} \notin \{-1,1\}^N$, then we choose an index $j$ with $-1 < [\bx_{\sf LP}]_{j} < 1$ and branch on variable $x_{j}$ by partitioning problem $(\mathcal{F}_+, \mathcal{F}_-, \setS)$  into two new subproblems $(\mathcal{F}_+ \cup \{j\}, \mathcal{F}_-, \setS)$ and $(\mathcal{F}_+ , \mathcal{F}_-\cup \{j\}, \setS)$.
	We add the two subproblems into the problem set $\mathcal{P}$.
\end{itemize}
The above process is repeated until $\mathcal{P}=\varnothing$.
The whole procedure is summarized as Algorithm \ref{Alg:OTF2}.

\begin{algorithm}[t!]
	\caption{A Global Algorithm for Solving Problem \eqref{eq:ML3}}\label{Alg:OTF2}
	\begin{algorithmic}[1]
		\renewcommand{\algorithmiccomment}[1]{~~//\,\texttt{#1}}
		\STATE {\bf input:} Initialize $\mathcal{P}=\{ (\varnothing, \varnothing, \setS) \}$ for   some $\setS \subseteq \setC$ and $U\leftarrow +\infty$.
		\WHILE{$\mathcal{P}\neq\varnothing$}
		\STATE Choose a subproblem $(\mathcal{F}_+, \mathcal{F}_-,\setS) \in \mathcal{P}$ and set $\mathcal{P} \leftarrow \mathcal{P}\backslash \{(\mathcal{F}_+, \mathcal{F}_-,\setS)\}$;
		\LOOP
		\STATE Solve the LP problem \eqref{eq:rel2} to obtain its optimal solution $(\bx_{\sf LP}, \bw_{\sf LP})$ and objective value $f_{\sf LP} = \sum_{i=1}^N [\bw_{\sf LP}]_i $;
		\IF{${f}_{\sf LP}  \geq U$}
			\STATE {\bf break};{\COMMENT{Corresponding to case {(i)}}}
		\ELSIF{$\bx_{\sf LP} \in \{-1,1\}^N$}
		\IF {$[\bw_{\sf LP}]_i \geq g_i (\bx_{\sf LP} )$ for all $i=1,2,\ldots,M$}
			\STATE Update $(\check{\bx},\check{\bw}) \leftarrow (\bx_{\sf LP}, \bw_{\sf LP})$ and $U \leftarrow {f}_{\sf LP}$;
			\STATE {\bf break}; {\COMMENT{Corresponding to case {(ii.1)}}}
		\ELSE
			\STATE $\setS \leftarrow\setS \cup  \{(i, \bx_{\sf LP}) \mid  [\bw_{\sf LP}]_i < g_{i}(\bx_{\sf LP}),  i=1,2,\ldots, M \}$; \COMMENT{Corresponding to case {(ii.2)}}
		\ENDIF
		\ELSE
			\STATE Choose an index $j$ such that $-1 < [\bx_{\sf LP}]_{j} < 1$;
\STATE  Add two new subproblems $(\mathcal{F}_+ \cup \{j\}, \mathcal{F}_-, \setS)$ and $ (\mathcal{F}_+ , \mathcal{F}_-\cup \{j\}, \setS)$ into $\mathcal{P}$;~ {\COMMENT{Corresponding to case { (iii)}}}
			\STATE \bf break;
		\ENDIF
		\ENDLOOP
		\ENDWHILE
		\STATE {\bf output}  $(\bx^{\star}, \bw^{\star}) \leftarrow (\check{\bx},\check{\bw})$.
	\end{algorithmic}
\end{algorithm}

In lines 3 and 16 of Algorithm \ref{Alg:OTF2}, there exist different
strategies to choose a subproblem  $ (\mathcal{F}_+, \mathcal{F}_-,\setS)$ from set $\mathcal{P}$ and to choose a branching variable index $j$ \cite{Achterberg2009}.
It is  worth  remarking that Algorithm \ref{Alg:OTF2} can be embedded into state-of-the-art MILP solvers like CPLEX through the so-called \emph{callback routine} \cite{CPLEX2022}, which uses the (default) fine-tune subproblem selection and branching strategies of MILP solvers.

\section{An Amplitude Retrieval Approach}
\label{sec:AR}

Due to the one-bit quantization, the received signal $\br$ in~\eqref{eq:model} loses the amplitude information, which results in a more challenging likelihood function of the one-bit ML MIMO detection problem \eqref{eq:ML}  compared to that of the classic MIMO detection problem.
In this section, we go for a different direction by proposing to introduce the amplitude information into the detection formulation and present a new AR formulation as an alternative to the one-bit ML MIMO detection  \eqref{eq:ML}.
Compared with the ML formulation \eqref{eq:ML}, the proposed AR formulation has a much simpler objective function. This will bring new opportunities to build efficient algorithms that do not require to confront the complicated likelihood functions.

\subsection{AR Formulation}
\label{subsec:AR}
Observe that the probit model \eqref{eq:ML} keeps the sign information of $\by$ and thereby loses  the amplitude information.
We introduce a latent variable  $\bz\in \Rbb_{+}^M$ to denote the missing amplitude information associated with $\by$, i.e.,
\[
    \by = \br \odot  \bz,
\]
where $\odot$ is the elementwise product.
With the aid of $\bz$, a natural choice is to minimize the distance between $\by$ and $\bH \bx$.
Specifically, we consider  the following   problem formulation which minimizes the $\ell_1$ residual $\sum_{i=1}^{M}| r_i \cdot z_i - \bh_i^{\top} \bx|$:
\begin{equation}\label{eq:l1_min}
  \begin{split}
    \min_{\bx, \bz} &~\sum_{i=1}^{M}| r_i \cdot z_i - \bh_i^{\top} \bx|\\
    \mbox{s.t. }&~ \bx \in \{ -1,1 \}^N, ~ z_i \geq 0,  ~\forall~ i=1,2,\ldots, M.
  \end{split}
\end{equation}
By noting that $r_i\in \{ -1,1\}$ implies
\[
| r_i \cdot z_i - \bh_i^{\top} \bx|= | z_i -r_i \cdot\bh_i^{\top} \bx|,
\]
we can equivalently express  problem \eqref{eq:l1_min} into a compact form
\begin{equation}\label{eq:l1_min2}
  \begin{split}
    \min_{\bx, \bz} &~\| \bz - \bB \bx \|_1\\
    \mbox{s.t. }&~ \bx \in \{ -1,1 \}^N, ~ z_i \geq 0,  ~\forall~ i=1,2,\ldots, M,
  \end{split}
\end{equation}
where $\bB = \mbox{Diag}(\br) \bH$.

 Problem \eqref{eq:l1_min2} can be further simplified.
Given any $\bx$, the optimal $z_i$ to problem \eqref{eq:l1_min2} is given by
\begin{equation}\label{eq:z_opt}
  z_i = \max(\bb_i^{\top} \bx,0),\quad \forall~ i=1,2,\ldots, M.
\end{equation}
By substituting the solution \eqref{eq:z_opt}   into problem \eqref{eq:l1_min2}, problem \eqref{eq:l1_min2} can be cast into
\begin{equation}\label{eq:AR}
  \begin{split}
    \min_{\bx} &~\sum_{i=1}^{M} \max(- \bb_i^{\top} \bx,   0)  =  \| \max(- \bB \bx, \bm 0)  \|_1\\
     \mbox{s.t. } &~\bx \in \{ -1,1 \}^N.
  \end{split}
\end{equation}
The above AR formulation provides an alternative handle to  the one-bit ML MIMO detection formulation in \eqref{eq:ML}.
Below let
 us compare the two formulations and reveal more insights.


First, we take a closer look at the term $\bb_i^{\top} \bx$ (or $r_i \cdot \bh_i^{\top} \bx$).
Suppose $\bb_i^{\top} \bx >0$, which means $y_i$ and the noiseless signal $\bh_i^{\top}\bx$ have the same sign.
Intuitively, this is likely to be true since the noise power $\sigma^2$ is usually small compared to the power of signal $\bH\bx$.
It is seen that the objective function in \eqref{eq:AR} endeavors to encourage $\bb_i^{\top} \bx $ for all $i$ being positive.
In fact, the  ML formulation \eqref{eq:ML} has a similar effect, i.e., minimizing the $-\log \Phi$ function also pushes $\bb_i^{\top} \bx $ for all $i$ to be positive.
If one replaces  $\max(-x,0)$ by $-\log \Phi(x)$ in problem \eqref{eq:AR}, it leads to
\begin{equation}\label{eq:AR_ML}
\begin{split}
\min_{\bx} &~- \sum_{i=1}^M \log \Phi(\bb_i^{\top} \bx)  \\
 \mbox{s.t. }& \bx \in \{-1 ,1\}^N.
 \end{split}
\end{equation}
We see that problem  \eqref{eq:AR_ML} is very similar to the one-bit ML MIMO detection problem \eqref{eq:ML}, except that problem  \eqref{eq:AR_ML} does not involve the noise standard variance $\sigma$.
This revelation connects the AR formulation with the ML problem, albeit the former is originally developed from a different rationale from ML.

Second, the design with the AR rationale can be flexible.
Instead of considering the formulation \eqref{eq:l1_min}, one can also consider minimizing other possible loss functions.
For example, one can  minimize the squared residue, which is given by
\begin{equation}\label{eq:l2_min}
  \begin{split}
    \min_{\bx, \bz} &~\sum_{i=1}^{M}| r_i \cdot z_i - \bh_i^{\top} \bx|^2\\
    \mbox{s.t. }&~ \bx \in \{ -1,1 \}^N, ~ z_i \geq 0,  ~\forall~ i=1,2,\ldots, M.
  \end{split}
\end{equation}
In the same vein of the development in \eqref{eq:z_opt}{--}\eqref{eq:AR}, problem \eqref{eq:l2_min} can be cast into
\begin{equation}\label{eq:AR2}
  \begin{split}
    \min_{\bx}&~ \| \max(- \bB \bx, \bm 0)  \|_2^2\\
     \mbox{s.t.}&~ \bx \in \{ -1,1 \}^N.
  \end{split}
\end{equation}
It is seen that the objective function in \eqref{eq:AR2} is quadratic in $\bx$, and is simpler than that of one-bit ML MIMO detection problem \eqref{eq:ML}.
As  discussed before, the probability  $\bb_i^{\top} \bx <0$ is likely to be low, i.e., only for a small portion of the indices $ \{ 1, 2,\ldots, M \}$ the cases $\bb_i^{\top} \bx <0$ can happen.
This can be understood from a sparsity pursuit perspective.
Motivated by this, we prefer the formulation \eqref{eq:AR} to \eqref{eq:AR2}, because the $\ell_1$-norm has a better capability of pursuing sparsity than the $\ell_2$-norm.
We will provide simulations to verify this in Section~\ref{sec:sim}.

We should mention related works.
The  AR idea was  exploited under a different context,  namely, channel estimation for Frobenius norm minimization \cite{qian2019amplitude}.
The resulting formulation and algorithm there are different from those of this paper.
In addition, in the context of nonnegative matrix factorization, our previous work \cite{huang2022sisal} utilized  $-\log \Phi(x)$, $\max(-x,0)$,  and $\max(-x,0)^2$  as penalty functions for promoting nonnegativity.
 However, it was not derived from the AR idea.

Third, the objective function of the AR problem \eqref{eq:AR} is piecewise linear and convex.
This simple form enables to design efficient algorithms that do not require confronting the challenging likelihood function  in the one-bit ML MIMO detection problem.
In particular, we rewrite problem \eqref{eq:AR} into the following standard MILP problem
\begin{equation}\label{eq:ARlinear}
  \begin{split}
    \min_{\bx, \bw} &~\sum_{i=1}^{M} w_i \\
     \mbox{s.t. } &~w_i \geq -\bb_i^\top \bx,~\forall~i =1,2, \ldots, M,\\
     & ~\bx \in \{ -1,1 \}^N,~\bw \in \mathbb{R}_+^M,
  \end{split}
\end{equation}
where we introduce a nonnegative variable $\bw$ to play the role of $\max(\cdot, 0)$ in problem \eqref{eq:AR}.
In sharp contrast to the one-bit ML MIMO detection problem \eqref{eq:ML}, the AR problem \eqref{eq:AR} (or its equivalent MILP form \eqref{eq:ARlinear}) can be directly solved to global optimality using off-the-shelf MILP solvers such as CPLEX.

Finally, we present the complexity result of problem \eqref{eq:AR}.
Although the AR problem \eqref{eq:AR} is much easier to handle than the one-bit ML MIMO detection problem \eqref{eq:ML}, it is also   NP-hard, as detailed in the following theorem.
\begin{Theorem}\label{thm:AR_NP1}
The AR problem \eqref{eq:AR} is NP-hard.
\end{Theorem}

\noindent{\it Proof}: The proof is similar to that of Theorem \ref{thm:ML_NP}.
In particular, we prove that  there exists a special instance of problem \eqref{eq:AR} which is as hard as the  partition problem.
	
Given any instance of the partition problem, we construct an instance of problem \eqref{eq:AR} by setting  $M=2$, $N=n$, $\sigma=1$, $\bb_1= \bw$ and $\bb_2 = -\bw$.
By construction, problem \eqref{eq:AR} reduces to
\begin{equation}\label{eq:MLxex1}
   { v^{\ast}} = \min_{\bx\in \{ -1,1 \}^n} \left(\max\left\{ - \bw^\top\bx, 0\right\} + \max\left\{ \bw^\top \bx, 0\right\}\right)
\end{equation}
For any $\bx \in \{-1,1\}^n$, we have $\max\{-\bw^\top \bx, 0\}\geq 0$ and $\max\{ \bw^\top \bx, 0\}\geq 0$, implying that $ v^{\ast} \geq 0$.
Moreover, $ v^{\ast} = 0$ if and only if $\bw^\top \bx=0$ holds for some $\bx \in \{-1,1\}^n $.
Similarly, the latter is equivalent to that the answer to the partition problem is yes, and hence problem \eqref{eq:AR} is NP-hard.~\hfill$\blacksquare$

\medskip

In order to strike a better balance between the detection performance and the computational complexity, we will propose an efficient first-order algorithm to solve problem \eqref{eq:AR} in the next subsection.

%



%

\subsection{Alternate Barzilai-Borwein  Method for AR Problem \eqref{eq:AR}}


The algorithmic development in this subsection is based on the penalty technique and the projected gradient (PG) method.
 More specifically, we  first transform problem \eqref{eq:AR} into a  convex box-constrained smooth problem by a smoothing technique and a penalty method;
 then, we apply a modified PG method to solve the transformed problem.
The resulting algorithm has a low per-iteration complexity.

We begin with a more generic optimization problem of the form:
\begin{equation}\label{eq:opt}
  \begin{split}
    \min_{\bx} &~\sum_{i=1}^{M}\max \{ \ba_{i,1}^{\top} \bx, \ba_{i,2}^{\top}\bx \} \\
     \mbox{s.t. }&~ \bx \in \{-1,1\}^N,
  \end{split}
\end{equation}
where  $\ba_{i,k}\in \Rbb^{N}$ for all $(i,k)$.
 The AR problem \eqref{eq:AR} is a special case of  problem \eqref{eq:opt} with   $\ba_{i,1} = - \bb_{i}  $ and $\ba_{i,2} = \bm 0$ for all $i$.

Problem \eqref{eq:opt} has two main challenges if we would like to apply first-order algorithms such as the PG algorithm  for solving it: (i) the objective function is nonsmooth;
(ii) the constraint is binary.
We tackle the nonsmoothness by applying the smoothing technique in \cite{nesterov2005smooth}.
To describe, we formulate problem \eqref{eq:opt}  as a min-max problem as follows:
\begin{equation}\label{eq:opt2}
  \begin{split}
    \min_{\bx} &~ \max_{\{\bm\theta_i\}_{i=1}^M} \sum_{i=1}^{M} \langle\bA_i \bx, \bm\theta_i \rangle  \\
     \mbox{s.t. }&~ \bx \in \{-1,1\}^N,~ \bm \theta_i \in \Delta_2,  ~\forall~ i=1,2,\ldots, M,
  \end{split}
\end{equation}
where $\bA_i = [\ba_{i,1},~\ba_{i,2}]^{\top}$ and
\[
\Delta_2 : = \{ \bm \theta \in \Rbb^2\mid \theta_1+ \theta_2=1, \theta_1, \theta_2 \geq 0 \}
\]
 is the two-dimensional unit simplex.
The smoothing approximation of problem \eqref{eq:opt2} is given by
\begin{equation}\label{eq:opt3}
  \begin{split}
    \min_{\bx} &~\varphi(\bx)    \\
     \mbox{s.t. }&~ \bx \in \{-1,1\}^N,~\quad
  \end{split}
\end{equation}
where
\begin{equation}\label{eq:smooth}
  \begin{split}
\varphi(\bx): = &\max_{\{\bm\theta_i\}_{i=1}^M} {\sum_{i=1}^{M}\left(\langle\bA_i \bx, \bm\theta_i \rangle -\frac{\rho}{2} \|\bm \theta_i \|_2^2\right)}\\
     &~\mbox{s.t. } \bm \theta_i \in \Delta_2,  ~ i=1,2,\ldots, M,\\
  \end{split}
\end{equation}
for a given smoothing parameter $\rho > 0$.
Obviously, problem \eqref{eq:smooth} reduces to problem \eqref{eq:opt2} when $\rho =0$.
In problem \eqref{eq:opt3},
adding the term $-\frac{\rho}{2} \|\bm \theta_i \|_2^2$ makes the function $\varphi(\bx)$  smooth~\cite{nesterov2005smooth}.
As a consequence, the gradient of $\varphi(\bx)$ is given by
\begin{equation}\label{eq:grad}
    \nabla \varphi(\bx) = \sum_{i=1}^{M} \langle\bA_i \bx, \bm\theta_i^{\ast}(\bx) \rangle,
\end{equation}
where $\bm\theta_i^{\ast}(\bx)$ is the unique optimal solution of the inner problem within $\varphi(\bx)$ at a fixed $\bx$, which can be obtained in a closed form,  i.e.,
\begin{equation*}
\begin{split}
 \theta_{i,1}^{\star}(\bx) =&~ \left[\frac{\rho +(\ba_{i,1}- \ba_{i,2})^{\top}\bx}{2\rho} \right]_{0}^1,\\
 \theta_{i,2}^{\star}(\bx) = &~ 1- \theta_{i,1}^{\star}(\bx),
 \end{split}
\end{equation*}
with $[x]_{a}^b: = \max\{a,\min\{x, b\}\}$ defined as the projection operator for projecting $x$ onto the  interval $[a,b]$.
By \cite[Theorem 1]{nesterov2005smooth},  the gradient $\nabla \varphi(\bx)$ is $L_{\varphi}$-Lipschitz continuous, i.e.,
\[
    \| \nabla \varphi(\bx_1) -  \nabla \varphi(\bx_2) \|_2 \leq L_{\varphi} \| \bx_1 - \bx_2 \|_2,
\]
for any $\bx_1,\bx_2 \in \Rbb^{N}$,
where
\[
L_{\varphi} = \frac{1}{\rho}\sum_{i=1}^{M} \sigma_{\max}(\bA_i)^2
\]
with $\sigma_{\max}(\bA)$ denoting the maximum singular value of $\bA$.

Next, we tackle the
 binary constraint  by the extreme-point pursuit penalty method \cite{shao2019framework,liu2024extreme}.
Specifically, we further reformulate problem \eqref{eq:opt3} as follows:
\begin{equation}\label{eq:EXPP}
  \begin{split}
    \min_{\bx} &~F_{\lambda}(\bx):=\varphi(\bx)  - \lambda \| \bx \|_2^2\\
     \mbox{s.t. } &~\bx \in [ -1,1 ]^N,
  \end{split}
\end{equation}
where $\lambda\geq 0$ is a penalty parameter.
The crux is to relax the binary constraint $\{-1,1\}^N$  as the box constraint $[-1,1]^N$,  and to use the negative square penalty $-\lambda\|\bx \|_2^2$ to force the solution to be an extreme point of $[-1,1]^N$, i.e., a point in $\{-1,1\}^N$.
Based on the result in \cite{shao2019framework,liu2024extreme}, as long as $\lambda > L_{\varphi}/2$, problems \eqref{eq:opt3} and \eqref{eq:EXPP} have the same optimal solutions.
The above transformation turns the discrete constraint into a continuous one, thereby allowing us to take advantage of the rich continuous optimization techniques for the algorithmic design.

Now we are ready to present the first-order algorithm for tackling problem \eqref{eq:EXPP}.
We apply the ABB method \cite{dai2005projected}.
The ABB method is a PG method with modified step sizes, which is motivated by
Newton’s method but does not involve any Hessian.
To be specific, the step size  of the ABB method in the $\ell$th iteration is given by
\begin{equation}\label{eq:step_size}
       \alpha_{\ell} = \begin{cases}
                      \frac{\|\bs_{\ell}\|_2^2 }{|\langle\bs_{\ell}, \bm\beta_{\ell}\rangle|} , & \mbox{if } \ell \mbox{ is even}; \\~\\[-0.2cm]
                      \frac{|\langle\bs_{\ell}, \bm\beta_{\ell}\rangle|}{\|  \bm\beta_{\ell}\|_2^2} , & \mbox{if } \ell \mbox{ is odd},
                    \end{cases}
\end{equation}
where
\begin{equation}\label{eq:s_beta}
  \begin{split}
     \bs_{\ell} =      \bx_{\ell}- \bx_{\ell-1},~
    \bm \beta_{\ell }  =   \nabla F_{\lambda}(\bx_{\ell}) -\nabla F_{\lambda}(\bx_{\ell-1}).
  \end{split}
\end{equation}
With the above step size, the ABB method performs the following  update
 \begin{equation*}
\begin{split}
    \bar{\bx}_{\ell+1}  = &~ [\bx_{\ell} - \alpha_{\ell} \nabla  F_{\lambda}(\bx_{\ell} ) ]_{-\bm 1}^{\bm 1}.
    \end{split}
\end{equation*}
Then, the ABB method searches the next iterate along the direction
\[
\bd_{\ell} = \bar{\bx}_{\ell+1} - \bx_{\ell}.
\]
We choose the Grippo-Lampariello-Lucidi (GLL) line search~\cite{grippo1986nonmonotone}, which,
given a chosen $\tau\in (0,1)$, finds $\eta>0$ decreasing from $1$ such that
\begin{equation}\label{eq:GLL}
  \begin{split}
    F_{\lambda}(\bx_{\ell} + \lambda \bd_{\ell}) \leq f_r + \tau \eta \langle\nabla F_{\lambda}(\bx_{\ell}), \bd_{\ell} \rangle,\\
    f_r :=\max \{ F_{\lambda}(\bx_{\ell-i}) : 0\leq i\leq \kappa-1\}.
  \end{split}
\end{equation}
Note that the objective function sequence $\{F_{\lambda}(\bx_{\ell})\}$ obtained by the GLL line search can be nonmonotone.
This may help avoid the solution getting stuck at bad local minima~\cite{dai2005projected}.

We apply the above ABB method to solve problem \eqref{eq:EXPP}, of which the detailed implementation is shown in Algorithm~\ref{Alg:ABB}.
Note that in the algorithm, we apply homotopy optimization, which is an optimization paradigm that traces a path from the solution of an easy problem to that of the target problem by the use of a homotopy.
By doing so, the path-tracing strategy  may effectively avoid bad local minima,  as numerical results suggest.
It is worth noting that  homotopy optimization was successfully applied in many signal processing and machine learning applications, demonstrating promising performance;
the readers are referred to \cite{xiao2013proximal,shao2020binary} and the references therein for more details of homotopy optimization.
Here, the transformation is problem \eqref{eq:EXPP} and $\lambda$ serves as the homotopy parameter.
Algorithm~\ref{Alg:ABB} starts from a small $\lambda$ and gradually increases its value.
When $\lambda =0$, problem \eqref{eq:EXPP} is a convex relaxation of problem \eqref{eq:AR} and is easy to solve; when $\lambda$ is large, problem \eqref{eq:EXPP} becomes closer to problem \eqref{eq:AR}.
%

\begin{algorithm}[t!]
\caption{The ABB Method for Problem \eqref{eq:EXPP}}\label{Alg:ABB}
\begin{algorithmic}[1]
\STATE Initialize $\bx_{-1}$, $\bx_0$, $\lambda$, $\lambda_{\max}$, $\rho$,  $\tau\in (0,1)$, $\kappa$, and $c$; $\ell=0$.
\WHILE{$\lambda<\lambda_{\max}$}
\REPEAT
\STATE update $\ell \leftarrow \ell+1$
\STATE calculate
\[
    \bs_{\ell} \leftarrow    \bx_{\ell}- \bx_{\ell-1}
                     , \quad
    \bm \beta_{\ell }  \leftarrow \nabla F_{\lambda}(\bx_{\ell}) -\nabla F_{\lambda}(\bx_{\ell-1});
\]

\STATE  calculate the ABB step size according to \eqref{eq:step_size};

\STATE update $\bm \theta_i$'s according to
\[
    \theta_{i,1}^{\ell} \leftarrow \left[ \frac{\rho - \bb_{i}^{\top}\bx_{\ell} }{2\rho}\right]_{0}^{1}, \quad ~\forall~ i=1,2,\ldots, M;
\]

\STATE  calculate the gradient
\begin{equation*}
  \begin{split}
    \nabla F_{\lambda}(\bx_{\ell}) \leftarrow  -\sum_{i=1}^{M} \theta_{i,1}^{\ell} \bb_{i}  - 2 \lambda \bx_{\ell};
  \end{split}
\end{equation*}

\STATE calculate
\begin{equation*}
\begin{split}
    \bar{\bx}_{\ell+1} \leftarrow  &~[\bx_{\ell} - \alpha_{\ell} \nabla  F_{\lambda}(\bx_{\ell} ) ]_{-\bm 1}^{\bm 1};
    \end{split}
\end{equation*}

\STATE calculate the line search direction
\[
    \bd_{\ell}\leftarrow  \bar{\bx}_{\ell+1} - \bx_{\ell};
\]

\STATE apply  the GLL line search to get $\eta$ according to \eqref{eq:GLL};

\STATE set $\bx_{\ell+1} \leftarrow \bx_{\ell} + \eta \bd_{\ell}$;
\UNTIL {$\| \bx_{\ell+1} - \bx_{\ell}\|_2\leq \epsilon $ for some $\epsilon >0$.}
\STATE $\lambda \leftarrow \lambda\cdot c$
\ENDWHILE
\end{algorithmic}
\end{algorithm}

\section{Simulation Results}
\label{sec:sim}

This section presents simulation results of different algorithms for solving the one-bit MIMO detection problem.
First, we provide simulation results to compare  different formulations and provide insights.
Then, we test the  BER performance of the proposed methods and state-of-the-art designs.
Finally, we compare the computational complexity of different algorithms.

We generate the simulation data according to the complex model \eqref{eq:model_com} and then transform them into the real-valued data.
The channel matrix $\tilde{\bH}$ is generated by the element-wise i.i.d. complex circular Gaussian distribution with  zero mean and
unit variance.
The elements of $\tilde{\bx}$, the symbols, are  independently and identically distributed (i.i.d.) and drawn from the binary constellation $\{ \pm 1\pm \jj \}^{\tilde{N}}$.
The signal-to-noise ratio (SNR) is defined as
\[
    {\rm SNR} = \frac{\Exp[\| \tilde{\bH} \tilde{\bx} \|_2^2]}{\Exp[\| \tilde{\bv} \|_2^2]}.
    \]
    A total number of 5,000 Monte-Carlo trials were run to obtain the BER and runtimes of our proposed algorithms and the benchmarked algorithms.

We need to specify the settings for Algorithms~\ref{Alg:OTF2} and \ref{Alg:ABB}.
In Algorithm~\ref{Alg:OTF2}, we initialize $\setS$ as $\{ (i,\hat{\bx}) \mid i=1,2,\ldots, M\}$ where $\hat{\bx}$ is a quantized zero-forcing (ZF) solution, i.e., $\hat{\bx} = \mbox{sgn}({\bH}^{\dag}{\br})$.
We initialize Algorithm  \ref{Alg:ABB} by a Bussgang  theorem-inspired ZF scheme
\[
\bx_{-1} = \bx_0 =  \frac{\sqrt{\pi (N+ \sigma^2)}}{2} (\bH^{\top} \bH + \sigma^2 \bI)^{-1} \bH^{\top}  (\br - \bd),
\]
where $\bd$ is a random realization of ${\cal N}(\bm 0, (1-2/\pi) \bI)$; see~\cite{bussgang1952crosscorrelation}.
The parameter $\lambda$ is initialized as $\lambda = 0.1\cdot N$.
We set $\lambda_{\max} =100$ , $\rho = 0.3+ \log(1+\sigma)$, $\tau =0.1$, $\kappa =4$, and $c =5$.

\subsection{Validation of Different Formulations}

\begin{figure}[t!]
\centering
\begin{subfigure}{0.49\textwidth}
\includegraphics[width=\linewidth]{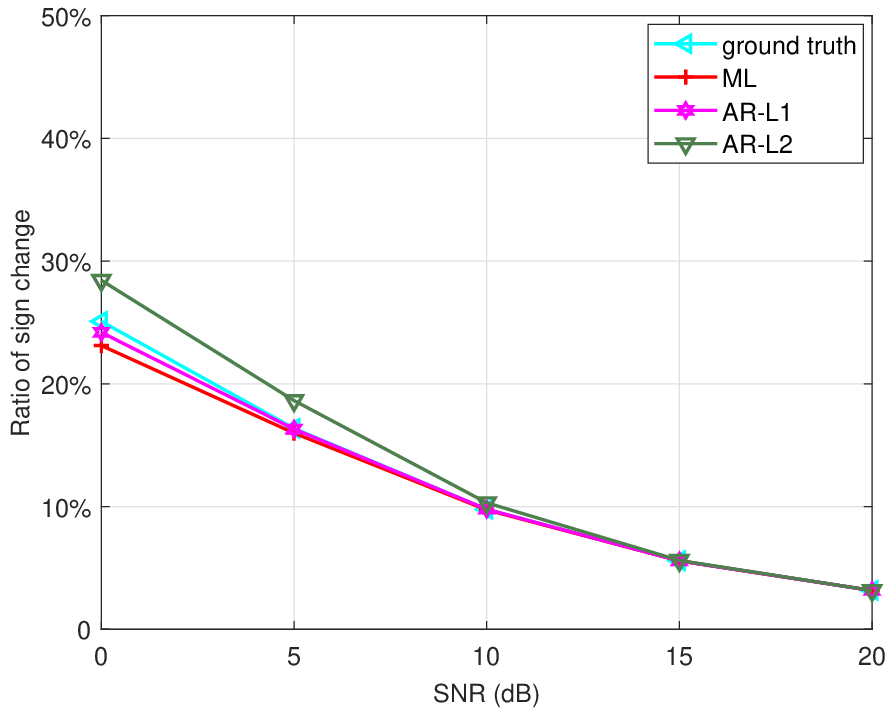}
\caption{$M=128$, $N=16$}
\end{subfigure}
\begin{subfigure}{0.49\textwidth}
\includegraphics[width= \linewidth]{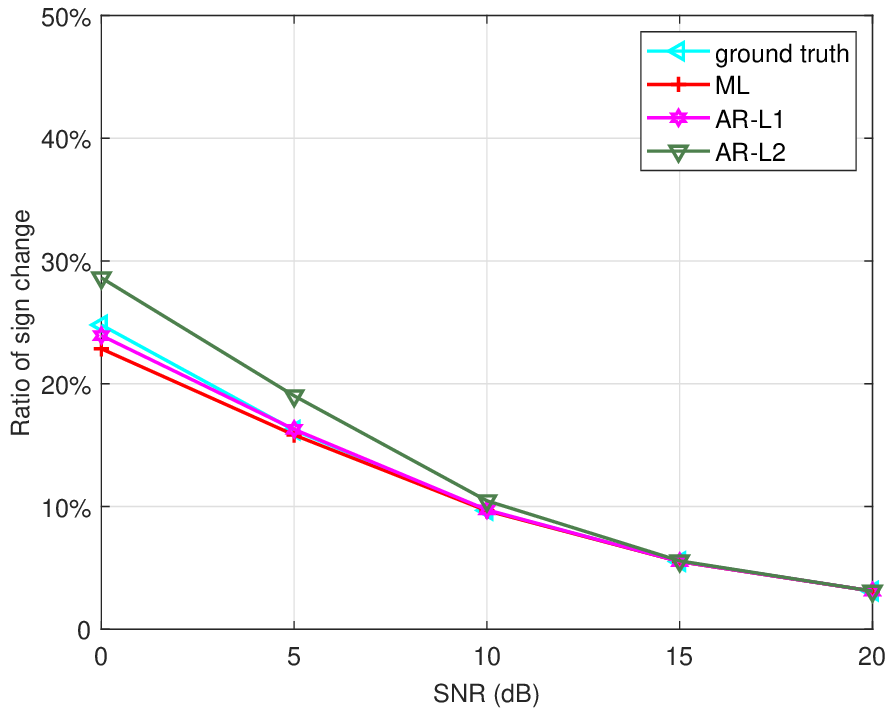}
\caption{$M=256$, $N=32$}
\end{subfigure}
\caption{The ratio between the total number of $i$ with $\bb_i^{\top} \bx<0$ and $M$.}
\label{fig:ratio}
\end{figure}

We first conduct a simulation to validate the effectiveness of the AR formulations.
Based on the discussion in Section \ref{sec:AR}, it is expected that the probability of $\bb_i^{\top} \bx<0$ is low, especially when the SNR is medium to high.
We verify this intuition by counting the ratio of $\bb_i^{\top} \bx<0$ for the solutions of different formulations; in other words, we would like to see in the signal model \eqref{eq:model}  how many elements of the received signal $\by$ have opposite signs with the noiseless signal $\bH \bx$.

In Fig. \ref{fig:ratio}, we consider (i) the  ground-truth transmit signal $\bx_{\sf GT}$;
(ii) the ML solution $\bx_{\sf ML}$ of problem \eqref{eq:ML};
(iii) the solution $\bx_{\sf AR\mh L1}$ of problem \eqref{eq:AR}, namely ``AR-L1'' in the legend;
and
(iv) the  solution $\bx_{\sf AR\mh L2}$  to the squared-loss AR variant \eqref{eq:AR2}, namely ``AR-L2'' in the legend.
The solutions in (ii)--(iv) are obtained by globally solving the corresponding problem formulations, respectively.

It can be seen from Fig.~\ref{fig:ratio} that the ground-truth transmit signal $\bx_{\sf GT}$ yields a low ratio of the sign change (between $y_i$ and $\bh_i^{\top} \bx_{ \sf GT}$), which coincides with the intuition that    noise can only affect a small portion of the signs, as  discussed in Section \ref{subsec:AR}.
Moreover, we  can observe from Fig.~\ref{fig:ratio} that the solutions to ML, AR-L1, and AR-L2 also yield comparably low ratios, which from this perspective validates the efficacy of ML and AR formulations.
In particular, when the SNR is low to medium, we see that both ML and AR-L1 are close to the ground truth, while AR-L2 has a higher ratio than the others.
The above observations suggest that the AR-L1  formulation may better approximate the ML formulation than AR-L2.

%

\subsection{BER Performance}

This subsection examines the BER performance of our proposed methods and state-of-the-art designs.
The considered methods are as follows:
\begin{enumerate}
  \item[(i)] quantized ZF detector\cite{choi2015quantized}, abbreviated herein as ``quant. ZF";
  \item [(ii)] ML problem \eqref{eq:ML}, globally solved by Algorithm~\ref{Alg:OTF2}, abbreviated  as  ``gML";
  \item [(iii)] ML problem \eqref{eq:ML}, a penalty method in our recent work~\cite{shao2020binary}, abbreviated  as  ``HOTML";
\item [ (iv)] AR problem \eqref{eq:AR}, globally solved by  calling CPLEX,  abbreviated  as ``AR-L1";
\item [ (v)] AR problem \eqref{eq:AR}, handled by Algorithm~\ref{Alg:ABB}, abbreviated  as ``AR-L1-ABB";
\item [(vi)] AR problem \eqref{eq:AR2} with the squared loss,  globally solved by calling CPLEX, abbreviated  as  ``AR-L2";
\item [ (vii)] near-maximum likelihood method,  abbreviated  as ``nML"~\cite{choi2016near};
\item [ (viii)] two-stage improved version of nML, abbreviated  as ``two-stage nML" \cite{choi2016near}.
\end{enumerate}

\begin{figure}[t!]
\centering
\begin{subfigure}{0.48\textwidth}
\includegraphics[width=\linewidth]{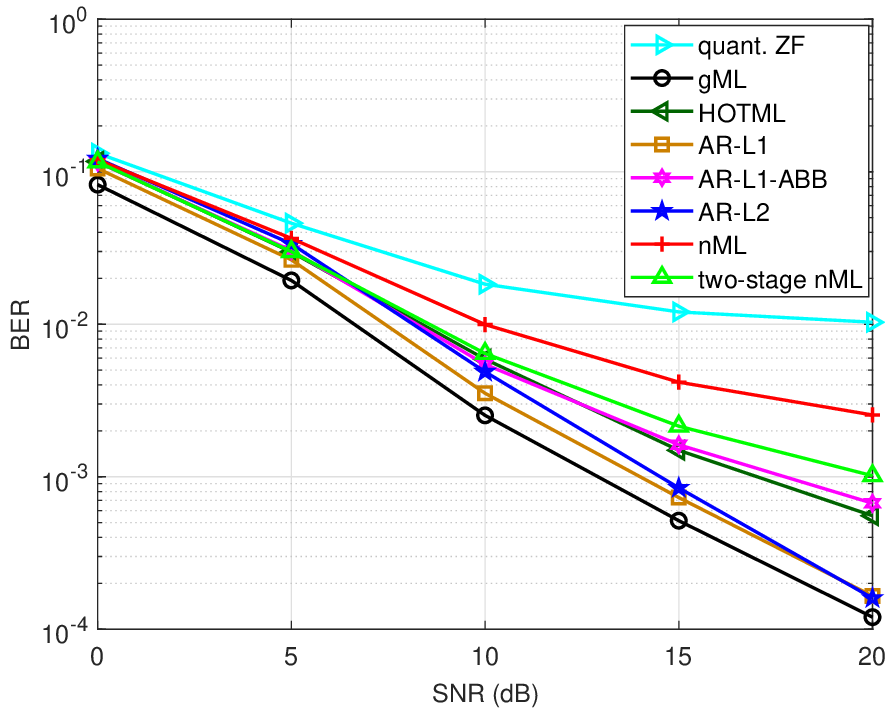}
\caption{$M=36$, $N=8$}
\end{subfigure}
\begin{subfigure}{0.48\textwidth}
\includegraphics[width= \linewidth]{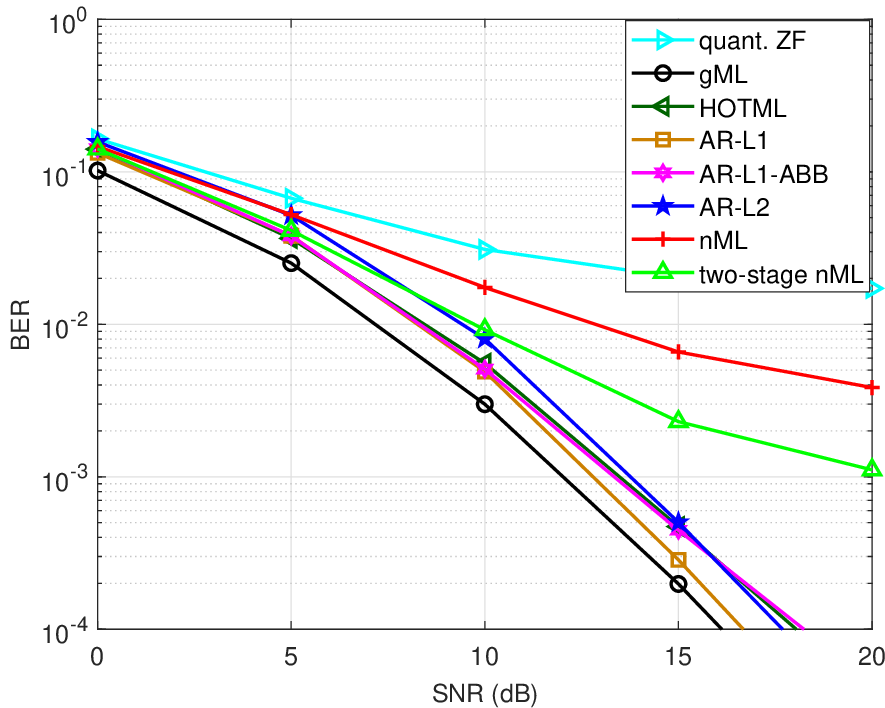}
\caption{$M=128$, $N=16$}
\end{subfigure}
\begin{subfigure}{0.48\textwidth}
\includegraphics[width= \linewidth]{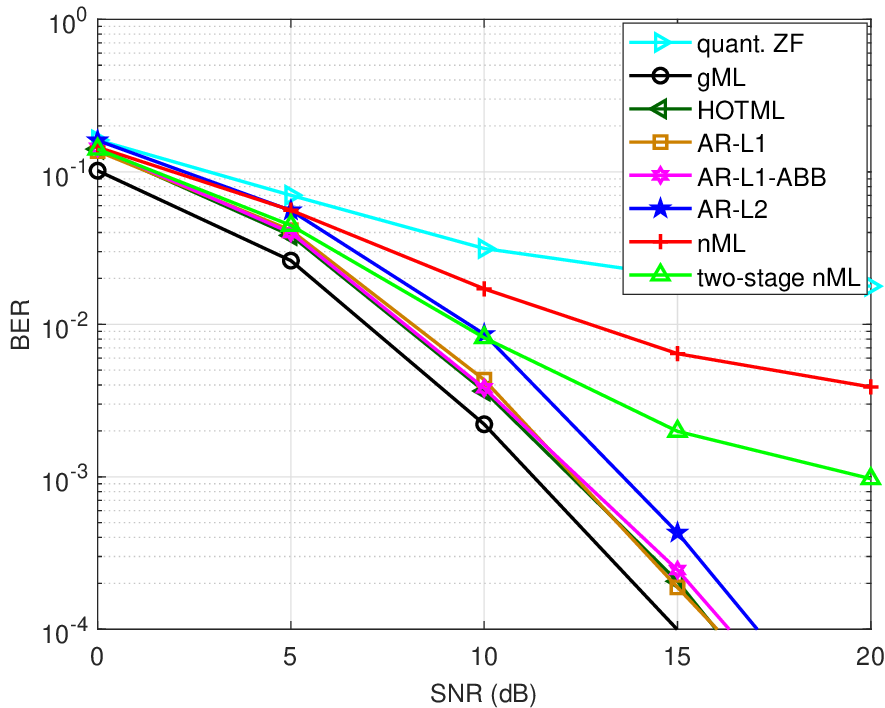}
\caption{$M=256$, $N=32$}
\end{subfigure}
\caption{BER performance of all compared approaches under different problem sizes.}
\label{fig:BER}
\end{figure}


The BER performance of  all  the above listed methods  under different system sizes is shown in Fig.~\ref{fig:BER}.
First, it is seen from the figure that  gML achieves the best BER performance, while  AR-L1 and AR-L2 follow with a slightly worse BER performance.
We can also see from Fig.~\ref{fig:BER} that AR-L1 performs better than AR-L2, which again  gives support to our intuition that the $\ell_1$-loss is  more preferred than the  $\ell_2$-loss in this problem.
Second, it  can be observed that the AR-L1-ABB method achieves a slightly worse BER performance than globally solving it by calling CPLEX, and the performance gap becomes narrow as the problem size increases.
In fact,  the AR-L1-ABB method achieves a comparable competitive performance with HOTML \cite{shao2020binary}.
In addition, both AR-L1-ABB method and HOTML perform much better than nML and its improved two-stage version.
We will see in the coming subsection that the promising performance of the AR-L1 method is achieved with a  low computational complexity.

\subsection{Runtime Comparison}

      \begin{figure}[t!]
\centering
\begin{subfigure}{0.49\textwidth}
\includegraphics[width=\linewidth]{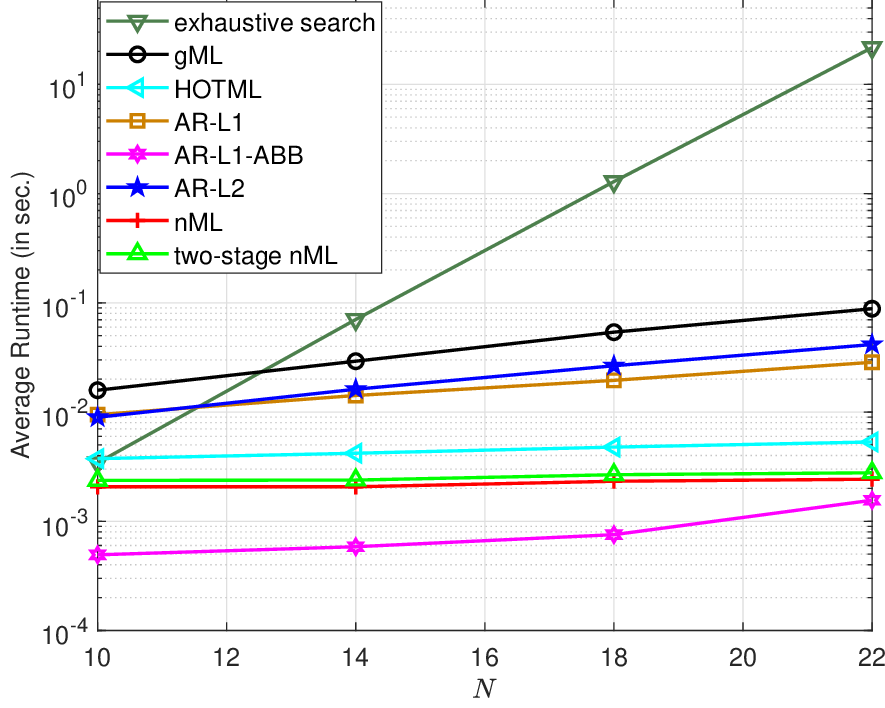}
\caption{$M/N =4$, SNR = $10$ dB}
\end{subfigure}
\begin{subfigure}{0.49\textwidth}
\includegraphics[width= \linewidth]{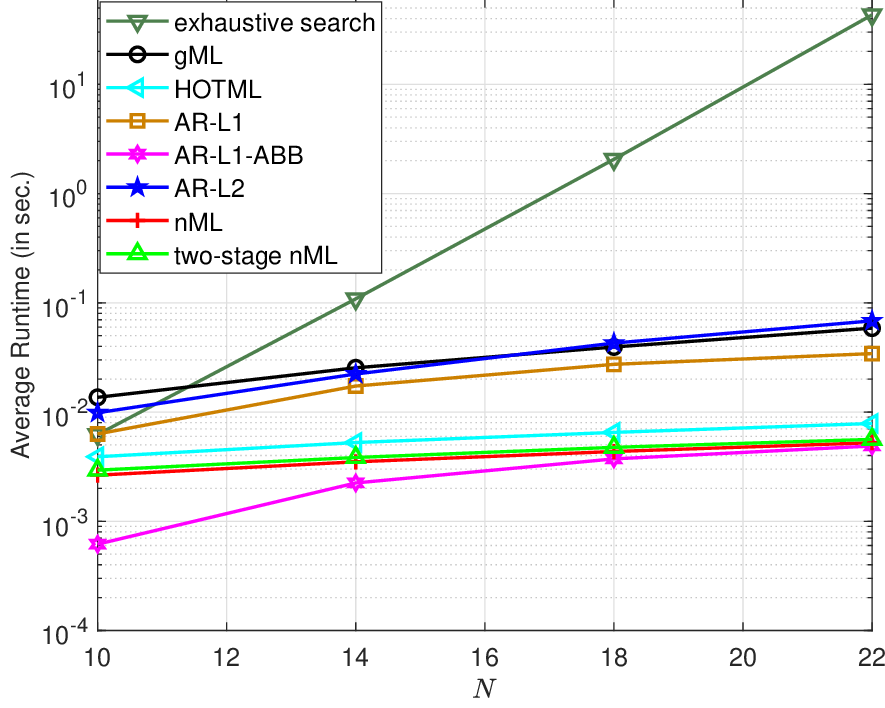}
\caption{$M/N =8$, SNR = $20$ dB}
\end{subfigure}
\caption{Runtime performance of all compared approaches with varying problem sizes.}
\label{fig:runtime1}
\end{figure}

In this subsection, we compare the runtime performance of the considered algorithms under different simulation settings.
First, we test how the average runtimes of the algorithms scale with the problem size.
The results are shown in Fig.~\ref{fig:runtime1}.
We increase $M$ and $N$ proportionally with fixed ratios $M/N$,  with the SNR being fixed.
From Fig.~\ref{fig:runtime1}, we see that the exhaustive search has a rapidly increasing runtime; when the problem size is large, it can be more than 100 times slower than the other algorithms.
By comparison, the runtime of gML scales much better than the exhaustive search.
It  can be seen from Fig.~\ref{fig:runtime1} that AR-L2 and AR-L1 are also computationally efficient.
 In particular, AR-L1 is the faster one between the two.
It is worth noting that the AR-L1-ABB method exhibits an attractively low running time, which is faster than HOTML.
This, combined with its outstanding BER performance, makes  AR-L1-ABB a competitive candidate algorithm.

\begin{figure}
  \centering
  \includegraphics[width=\linewidth]{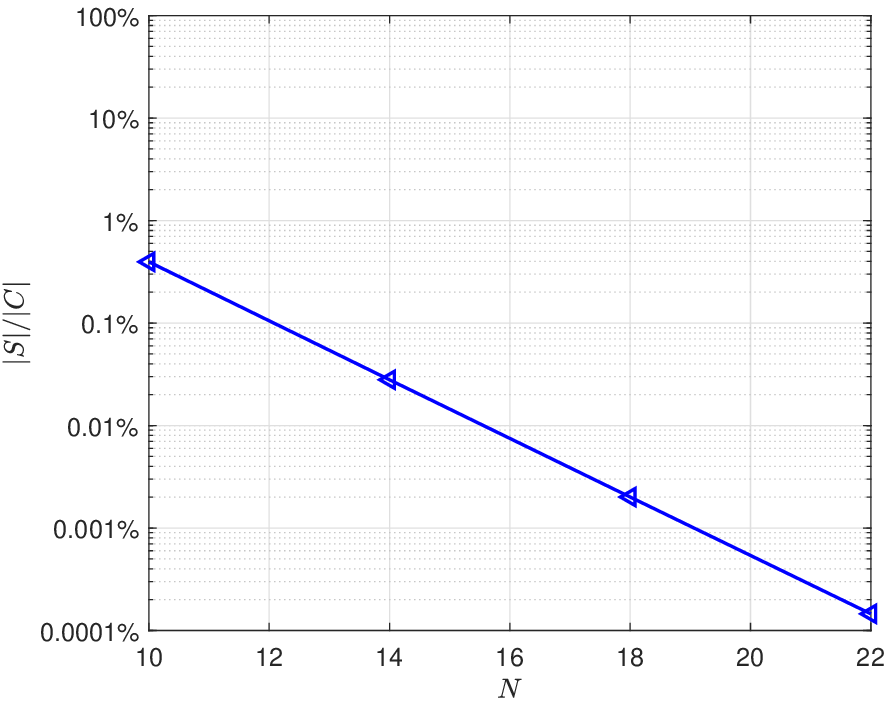}
  \caption{The ratio  $|\setS|/|\setC|$ when gML converges.}\label{fig:active_cons}
\end{figure}

To get a better idea with the efficiency of gML, we show the ratio $|\setS|/|\setC|$ when gML converges and outputs a global optimal solution to the ML problem \eqref{eq:ML}.
The result is shown in Fig.~\ref{fig:active_cons}.
We set $M=256$ and SNR$=10$ dB, and vary $N$.
It can be seen from Fig.~\ref{fig:active_cons} that the ratio $|\setS|/|\setC|$ is below $1\%$, which means that gML only needs to solve LP relaxation subproblems that  involve less than $1\%$ linear inequality constraints than problem \eqref{eq:ML3}.
It is also encouraging to see that as the number of users increases, the ratio keeps decreasing rapidly.
This indicates that gML
has good scalability for massive systems with many users.

\begin{figure}[t!]
\centering
\begin{subfigure}{0.49\textwidth}
\includegraphics[width=\linewidth]{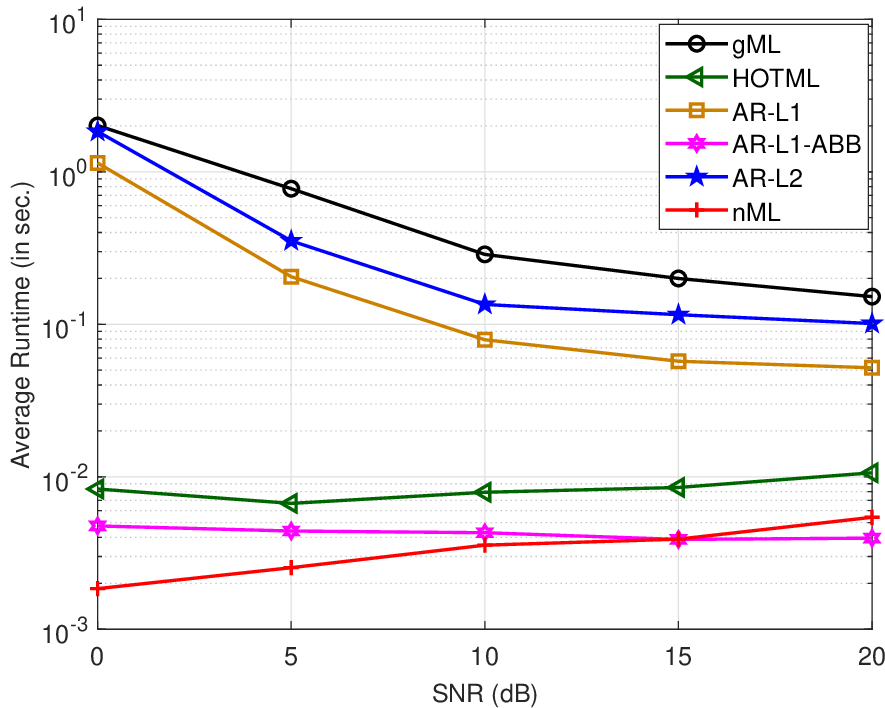}
\caption{$M=128$, $N=16$}
\end{subfigure}
\begin{subfigure}{0.49\textwidth}
\includegraphics[width= \linewidth]{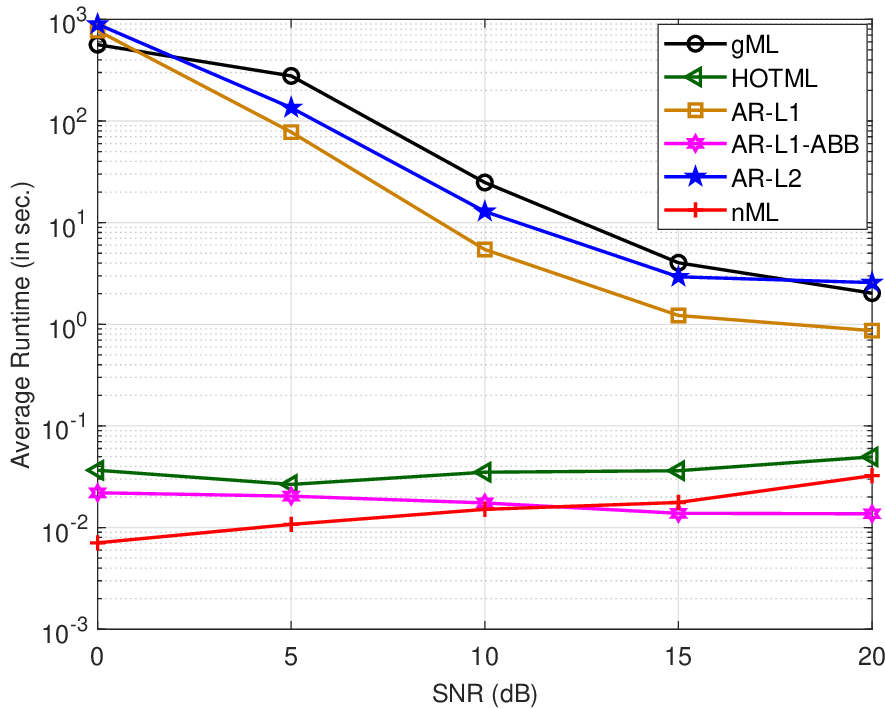}
\caption{$M=256$, $N=32$}
\end{subfigure}
\caption{Runtime performance of all compared algorithms with varying SNRs.}
\label{fig:runtime2}
\end{figure}

Fig.~\ref{fig:runtime2} shows the average runtimes of the considered algorithms under different SNRs with fixed problem sizes.
It is seen from the figure that the averaged runtimes of the  global algorithms, including gML, AR-L1 and AR-L2, decrease rapidly with the SNR.
On the other hand, the PG-based methods, including HOTML, nML and AR-L1-ABB, do not show such a significant decreasing runtime phenomenon.
In addition, it is seen from Fig.~\ref{fig:runtime2} that nML and HOTML, which tackle the one-bit ML MIMO problem,  show gradually increasing runtimes with the SNR.
Indeed, our recent work~\cite{shao2024accelerated} showed that the PG-based methods for handling the one-bit ML MIMO detection problem \eqref{eq:ML} may require more number of iterations to converge for high SNR regions.
By comparison, the AR-L1-ABB method handles the AR formulation \eqref{eq:AR} that has a different objective function  and does not have such issues.
Fig.~\ref{fig:runtime2} shows that  the runtime performance  of AR-L1-ABB does not grow with the SNR.

\section{Conclusion}
\label{sec:conc}

In this paper, we have proposed new formulations and efficient algorithms for the one-bit MIMO detection problem.
We first developed an efficient global algorithm for one-bit ML MIMO detection.
The proposed global algorithm is a customized branch-and-bound algorithm and can solve the  one-bit ML MIMO detection problem  with a significantly reduced runtime compared to the complete enumeration approach for large-sized problem instances.
The global algorithm provides an important performance benchmark  for performance evaluation of various existing  approximate algorithms for the same problem.
Moreover, we  have proposed a new AR formulation which introduces an explicit amplitude variable and minimizes the $\ell_1$-norm  residual between the reconstructed unquantized received signal and the noiseless  received signal.
The proposed AR formulation exhibits a simpler objective compared to that in the ML formulation.
 We have further devised an efficient first-order algorithm tailored  for the resulting AR formulation.
Our numerical results demonstrated that the proposed AR formulation can achieve a competitive BER performance while maintaining a low computational complexity, thereby striking a  balance between the BER performance and the runtime.



 \bibliographystyle{IEEEtran}
\bibliography{ref}

\end{document}